\documentclass[aps,prb,groupedaddress,twocolumn,showpacs]{revtex4-1}

\usepackage{graphicx}
\usepackage{epsfig}
\usepackage{color}
\usepackage{hyperref}
\usepackage{amssymb}

\usepackage{dcolumn}
\usepackage{bm}
\usepackage{amsfonts}
\usepackage{ulem}

\begin{document}

\newcommand\kk{{\bar{k}}}
\newcommand{\fab}[1]{{\bf \textcolor{red}{#1}}}


\title{Wigner-function formalism applied to semiconductor quantum devices: \\
Failure of the conventional boundary-condition scheme}


\author{Roberto Rosati${}^1$}

\author{Fabrizio Dolcini${}^{1,2}$}

\author{Rita Claudia Iotti${}^1$}
\author{Fausto Rossi${}^1$}
\email[]{Fausto.Rossi@polito.it}
\homepage[]{staff.polito.it/Fausto.Rossi}
\affiliation{${}^1$
Department of Applied Science and Technology, Politecnico di Torino, 
I-10129 Torino, Italy \\
${}^2$ CNR-SPIN, I-80126 Napoli, Italy}


\date{\today}

\begin{abstract}

The Wigner-function formalism is a well known approach to model charge transport in semiconductor nanodevices. Primary goal of the present article is to point out and explain intrinsic limitations of the conventional quantum-device modeling based on such Wigner-function paradigm, providing a definite answer to open questions related to the application of the conventional spatial boundary-condition scheme to the Wigner transport equation. Our  analysis shows that (i) in the absence of energy dissipation (coherent limit) the solution of the Wigner equation  equipped with given boundary conditions is not unique, and (ii) when decoherence/dissipation phenomena are taken into account via a relaxation-time approximation  the solution, although unique, is not necessarily a physical Wigner function.

\end{abstract}

 

\maketitle

\section{Introduction}\label{s-I}

Current micro/nanoelectronics technology pushes device dimensions toward space- and time-scales where the application of the traditional semiclassical or Boltzmann picture\cite{Jacoboni89,Jacoboni10} is questionable, and a comparison with genuine quantum approaches is highly desirable.\cite{Jacoboni10,Rossi11} 
However, in spite of the quantum-mechanical nature of electron and photon dynamics in the core region of typical solid-state nanodevices --e.g., superlattices\cite{Leo03} and quantum-dot structures\cite{Rossi05}-- the overall behavior of such quantum systems is often governed by a highly non-trivial interplay between phase coherence and dissipation/dephasing,\cite{Rossi02} the latter being also strongly influenced by the presence of space boundaries.\cite{Frensley90} \\
A widely used theoretical tool to account for such interplay  in  semiconductors is the  single-particle density matrix operator $\hat\rho$  for the electron subsystem.\cite{Rossi11,note-GF} The time evolution of $\hat{\rho}$ is given by the  density-matrix equation, which involves both a coherent-dynamics term and a scattering superoperator encoding the   energy-dissipative/decoherent interaction mechanisms that electrons experience within the host material.   The density-matrix  approach  applies to a variety of physical problems,\cite{Rossi11,Shah99} ranging from quantum-transport phenomena to ultrafast electro-optical processes in ``extended systems'', i.e., systems extending over the whole coordinate space. \\

\begin{figure}
\centering
\includegraphics*[width=7cm]{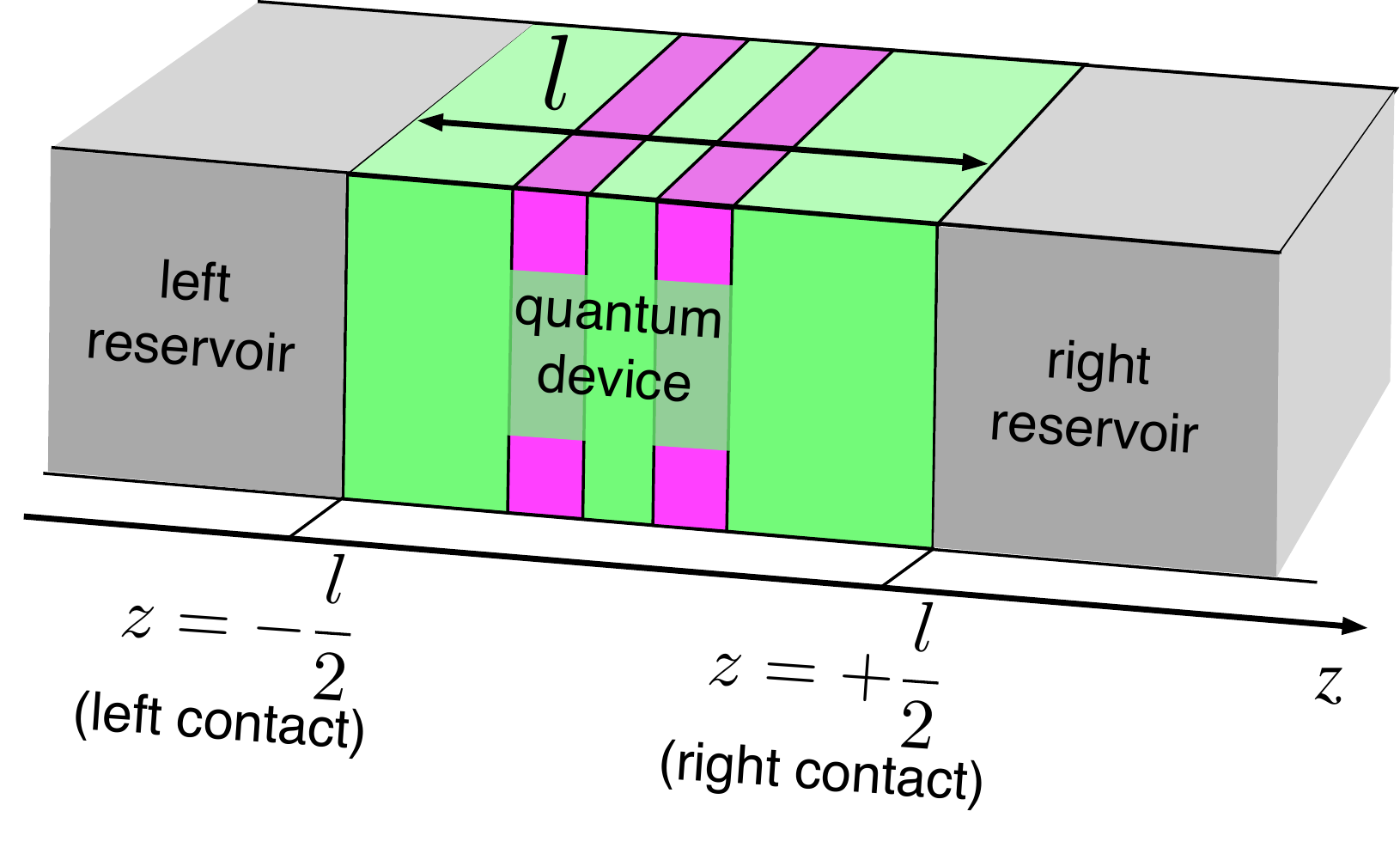}
\caption[]{(Color online) 
Schematic representation of a typical semiconductor-based quantum device as an open system connected to two external charge reservoirs. Here, the distance between the interfaces is $l$, and $z$ is the longitudinal transport direction.}
\label{fig1}       
\end{figure}

However, such approach cannot be straightforwardly applied to  nanostructured devices. Indeed, a typical nanodevice\cite{Rossi11} is  a ``localized system'', i.e., a portion of material characterized by a well defined volume and by spatial boundaries acting as electric contacts to external charge reservoirs, as sketched in Fig.~\ref{fig1}. Here,  $z$ denotes the transport direction, $l$ is the longitudinal length of the device,  the electric contacts being located at $z=-l/2$ and $z=+l/2$. The modeling of a nanostructure device  thus represents an intrinsically space-dependent problem, so that a real-space quantum treatment   accounting for the presence of quite different spatial regions becomes mandatory. 
To this purpose, the Wigner-function formalism\cite{Frensley90,Arnold08} is one of the adopted frameworks. 
Within this formalism, the statistical quantum state of the electronic subsystem is fully described in terms of the  Wigner function $f(\mathbf{r},\mathbf{k})$, 
defined over the conventional phase-space $(\mathbf{r},\mathbf{k})$ as the Weyl-Wigner
transform of the single-particle density-matrix operator $\hat{\rho}$.\cite{Toda83} \\

Based on the Wigner-function formalism, various approaches for the study of quantum-transport phenomena in semiconductor nanomaterials and nanodevices  have been proposed.\cite{Frensley86,Frensley87,Kriman87,Kluksdahl89,Buot90,Miller91,McLennan91,Tso91,Ferry93,Gullapalli94,Fernando95,Kim01,Nedjalkov04,Nedjalkov06,Querlioz08,Morandi09,Querlioz10,Yoder10,Jiang10,Jiang11,Savio11,Trovato11}
On the one hand, starting from the pioneering work by Frensley,\cite{Frensley86}  quantum-transport simulations based on a direct numerical solution of the Wigner equation have been performed via finite-difference approaches,\cite{Kluksdahl89} by imposing on the Wigner function the standard inflow or U boundary condition scheme (see Fig.~\ref{fig2} in Sec.~\ref{s-U}). On the other hand, a generalization to systems with open boundaries of the semiconductor Bloch equations has also been proposed.\cite{Rossi98c,Proietti03}
In addition to these two alternative simulation strategies --both based on effective treatments of relevant interaction mechanisms-- Jacoboni and co-workers have proposed a fully quantum-mechanical simulation scheme for the study of electron-phonon interaction based on the ``Wigner paths'';\cite{Pascoli98} this approach is intrinsically able to overcome the standard approximations of conventional quantum-transport models, namely the Markov approximation and the completed-collision limit;\cite{Rossi02} however, due to the huge amount of computation required, its applicability is often limited to short time-scales and extremely simplified situations.

Motivated by a few unphysical results\cite{Proietti03} obtained via the generalized semiconductor Bloch equations mentioned above, a recent study\cite{Taj06} has   shown that the application of the conventional inflow boundary-condition scheme to the Wigner transport equation may lead to partially negative charge probability densities, unambiguous proof of the failure of such classical-like Wigner-function treatment. \\

Primary goal of the present article is to point out that the  conventional boundary-condition scheme adopted for the solution of the Wigner equation exhibits some intrinsic limitations, whose impact may lead to totally unphysical results, especially in the coherent-transport regime. In order to illustrate this aspect, we shall mainly discuss a simple but physically relevant one-dimensional system, namely a delta-like potential profile. 
Our detailed analysis will show that (i) in the absence of energy dissipation (coherent limit) the solution of the Wigner equation (compatible with given boundary conditions) is not unique, and (ii) also when the solution is unique, the latter is not necessarily a Wigner function, i.e., a Weyl-Wigner transform of a single-particle density matrix. 

The article is organized as follows: In Sec.~\ref{s-FQDM} we shall summarize the fundamentals of quantum-device modeling, with a special focus on the problem of quantum systems with open space boundaries, corresponding, e.g., to the case of a semiconductor nanodevice inserted into an electric circuit. 
In Sec.~\ref{s-U} we shall discuss in very general terms the intrinsic limitations of the conventional boundary-condition scheme applied to a quantum-mechanical problem, thus addressing the main topic of the article, i.e., the physical versus unphysical nature of the Wigner-equation solutions corresponding to given spatial boundaries.
Section \ref{s-CL} is devoted to the investigation of the  coherent limit and of the corresponding Wigner equation, while Sec.~\ref{s-IERDP} deals with the inclusion of energy-dissipation and decoherence phenomena.
Finally, in Sec.~\ref{s-SC} we shall summarize and draw a few conclusions.

\section{Quantum-device modeling based on the Wigner-function formalism}\label{s-FQDM}

In order to account for the space-dependent character of a quantum device, a widely employed strategy is the Wigner-function treatment of the problem.\cite{Frensley90} The Wigner function $f(\mathbf{r},\mathbf{k})$ associated to a single-particle density-matrix operator $\hat{\rho}$ is defined as its Weyl-Wigner transform\cite{Toda83} 
\begin{eqnarray}\label{WF-op}
f(\mathbf{r},\mathbf{k}) &=&   \int d\mathbf{r}^\prime
e^{-{i \mathbf{k} \cdot \mathbf{r}'}}
\left\langle \mathbf{r} + {\mathbf{r}' \over 2}\right| \hat{\rho} \left|  \mathbf{r} - {\mathbf{r}' \over 2}  \right\rangle
\nonumber \\
&=& {\rm tr} \{ \hat{W}(\mathbf{r},\mathbf{k}) \hat{\rho} \}\ ,
\end{eqnarray}
corresponding to the quantum-plus-statistical average of the Wigner operator\cite{Rossi11}
\begin{equation}\label{hatW} 
\hat{W}(\mathbf{r}, \mathbf{k}) = \int d\mathbf{r}'
\left|\mathbf{r} - {\mathbf{r}' \over 2}\right\rangle
e^{-{i \mathbf{k} \cdot \mathbf{r}'}}
\left\langle \mathbf{r} + {\mathbf{r}' \over 2}\right|\ .
\end{equation}
In particular, for a  pure state $\vert\beta\rangle$ the  Wigner function  reduces to the expectation value of the Weyl-Wigner operator
\begin{equation}\label{WFps}
f_\beta(\mathbf{r},\mathbf{k}) = 
\langle \beta| \hat{W}(\mathbf{r},\mathbf{k}) | \beta \rangle\ .
\end{equation}
Within such Wigner-function representation the average values of charge and current densities at location $\mathbf{r}$ are given by
\begin{equation}
\label{dens-from-WF}
n(\mathbf{r}) = \int \frac{d\mathbf{k}}{(2\pi)^{3}} \, f(\mathbf{r},\mathbf{k}) 
\end{equation}
and
\begin{equation}
\label{cur-from-WF}
\mathbf{J}(\mathbf{r}) = \int \frac{d\mathbf{k}}{(2\pi)^{3}} \, \mathbf{v}(\mathbf{k}) \, f(\mathbf{r},\mathbf{k})\quad,
\end{equation}
where $\mathbf{v}(\mathbf{k})$ 
is the group velocity of an electron with wavevector $\mathbf{k}$. \\

The time evolution of the Wigner function can be derived from the equation of motion for the density-matrix operator:\cite{Rossi11}
\begin{equation}\label{SBE-op}
{d \hat{\rho}  \over dt} =  {1 \over i\hbar}\,\left[\hat H^\circ,  \, \hat{\rho}  \, \right] 
+ 
\Gamma \,(\hat{\rho})\ .
\end{equation}
Here, the first contribution on the r.h.s. describes the coherent dynamics dictated by a non-interacting Hamiltonian $\hat{H}^\circ$, including elastic single-electron scattering processes, while the second term is a linear superoperator~$\Gamma$ encoding the energy-dissipative/decoherent scattering mechanisms that electrons experience within the host material. 

By applying the Weyl-Wigner transform (\ref{WF-op}), together with its inverse
\begin{equation}\label{WF-op-inv}
\hat{\rho} = 
{1 \over (2\pi)^3}\,
\int d\mathbf{r} \int d\mathbf{k} \,
\hat{W}(\mathbf{r},\mathbf{k}) \,
f(\mathbf{r},\mathbf{k})  \, ,
\end{equation}
to the density-matrix equation~(\ref{SBE-op}), one obtains the equation of motion for the Wigner function
\begin{equation}\label{WE2}
{\partial f(\mathbf{r},\mathbf{k}) \over \partial t} = 
\left. {\partial f(\mathbf{r},\mathbf{k}) \over \partial t}\right|_\epsilon
+ 
\left. {\partial f(\mathbf{r},\mathbf{k}) \over \partial t}\right|_\Gamma
\end{equation}
with
\begin{equation}\label{WE2epsilon}
\left. {\partial f(\mathbf{r},\mathbf{k}) \over \partial t}\right|_\epsilon
= 
\int d\mathbf{r}^\prime \, d\mathbf{k}^\prime \epsilon(\mathbf{r},\mathbf{k};\mathbf{r}',\mathbf{k}') f(\mathbf{r}',\mathbf{k}')
\end{equation}
and
\begin{equation}\label{WE2Gamma}
\left. {\partial f(\mathbf{r},\mathbf{k}) \over \partial t}\right|_\Gamma
= 
\int  d\mathbf{r}^\prime \, d\mathbf{k}^\prime \, \Gamma(\mathbf{r},\mathbf{k};\mathbf{r}',\mathbf{k}') f(\mathbf{r}',\mathbf{k}')\ ,
\end{equation}
where 
\begin{equation}\label{epsilonWF}
\epsilon(\mathbf{r},\mathbf{k};\mathbf{r}',\mathbf{k}')
=
-{i \over (2\pi)^3 \hbar}\,{\rm tr}\left\{
\hat{W}(\mathbf{r},\mathbf{k})\,
\left[\hat H^\circ,\,\hat{W}(\mathbf{r}',\mathbf{k}')\right]
\right\}
\end{equation} 
and 
\begin{equation}\label{GammaWF}
\Gamma(\mathbf{r},\mathbf{k};\mathbf{r}',\mathbf{k}')
= {1 \over (2\pi)^3}\,
{\rm tr}\left\{
\hat{W}(\mathbf{r},\mathbf{k})\,\Gamma\left(\hat{W}(\mathbf{r}',\mathbf{k}')\right)
\right\}\end{equation} 
are the single-particle and the scattering superoperators written in the $(\mathbf{r}, \mathbf{k})$ Wigner picture, respectively. \\

For any given basis set $\{|\alpha\rangle\}$, the Wigner function (\ref{WF-op}) can also be expressed as  
\begin{equation}\label{WF}
f(\mathbf{r},\mathbf{k}) = \sum_{\alpha_1\alpha_2} W_{\alpha_2\alpha_1}(\mathbf{r},\mathbf{k}) \rho_{\alpha_1\alpha_2}
\end{equation}
where
\begin{equation}\label{calW}
W_{\alpha_2\alpha_1}(\mathbf{r},\mathbf{k}) = \int d\mathbf{r}^\prime
\phi_{\alpha_1}\left(\mathbf{r} + {\mathbf{r}' \over 2}\right)
e^{-{i \mathbf{k} \cdot \mathbf{r}'}}
\phi^*_{\alpha_2}\left(\mathbf{r} - {\mathbf{r}' \over 2}\right)\ ,
\end{equation}
with $\phi_\alpha(\mathbf{r}) = \langle \mathbf{r} \vert \alpha \rangle$ denoting the real-space wavefunction  corresponding to the basis state $|\alpha \rangle$.\\
In particular, by choosing as basis states $| \alpha \rangle$ the eigenstates of the noninteracting Hamiltonian 
\begin{equation}\label{hatHcirc}
\hat H^\circ = \sum_{\alpha} \vert\alpha\rangle \epsilon_{\alpha} \langle\alpha\vert \quad
\end{equation}
(corresponding to the energy spectrum $\epsilon_{\alpha}$), the single-particle density-matrix operator  can be expressed in terms of entries $\rho_{\alpha_1 \alpha_2}$ as
\begin{equation}\label{hatrho}
\hat\rho = \sum_{\alpha_1\alpha_2} \vert\alpha_1\rangle \rho_{\alpha_1\alpha_2} \langle\alpha_2\vert\quad,
\end{equation}
and the density-matrix equation (\ref{SBE-op}) is given by
\begin{equation}\label{SBE}
{d \rho_{\alpha_1\alpha_2} \over dt} = {\epsilon_{\alpha_1}-\epsilon_{\alpha_2}
\over i\hbar}  \rho_{\alpha_1\alpha_2} +\sum_{\alpha_1'\alpha_2'}
\Gamma_{\alpha_1\alpha_2, \alpha_1'\alpha_2'} \rho_{\alpha_1'\alpha_2'}\ .
\end{equation}
Such set of coupled equations of motion for the density-matrix elements $\rho_{\alpha_1\alpha_2}$ is usually referred to as the semiconductor Bloch equations.\cite{Rossi11}  

We  emphasize that in general the Wigner equation~(\ref{WE2}) is non-local  in both $\mathbf{r}$ and $\mathbf{k}$. As a consequence, the conventional boundary-condition scheme adopted to solve the semiclassical  Boltzmann equation cannot be applied.\cite{Rossi11}
However, in order to simplify the problem, the single-particle and scattering superoperators in Eqs.~(\ref{WE2epsilon}) and (\ref{WE2Gamma}) are often replaced by effective/phenomenological models. In particular, as we shall discuss in detail  in Sec.~\ref{s-CL}, within the conventional effective-mass and envelope-function approximations, the single-particle superoperator (\ref{WE2epsilon}) turns out to be local. 
Furthermore, adopting a generalized relaxation-time approximation (see also Sec.~\ref{s-IERDP}), the fully quantum-mechanical scattering superoperator~(\ref{WE2Gamma}) is replaced by the local form\cite{Frensley90}
\begin{equation}\label{RTAWF}
\left. {\partial f(\mathbf{r},\mathbf{k}) \over \partial t}\right|_\Gamma
= 
-\,{f(\mathbf{r}, \mathbf{k}) - f^\circ(\mathbf{r}, \mathbf{k}) \over \tau}\ ,
\end{equation}
describing the effect of dissipation/decoherence (induced by the host material) toward the equilibrium Wigner function $f^\circ(\mathbf{r}, \mathbf{k})$ in terms of a relaxation time $\tau$.

\section{The inflow or U boundary-condition scheme: mathematical versus physical solutions}\label{s-U}

\begin{figure}
\centering
\includegraphics*[width=8cm]{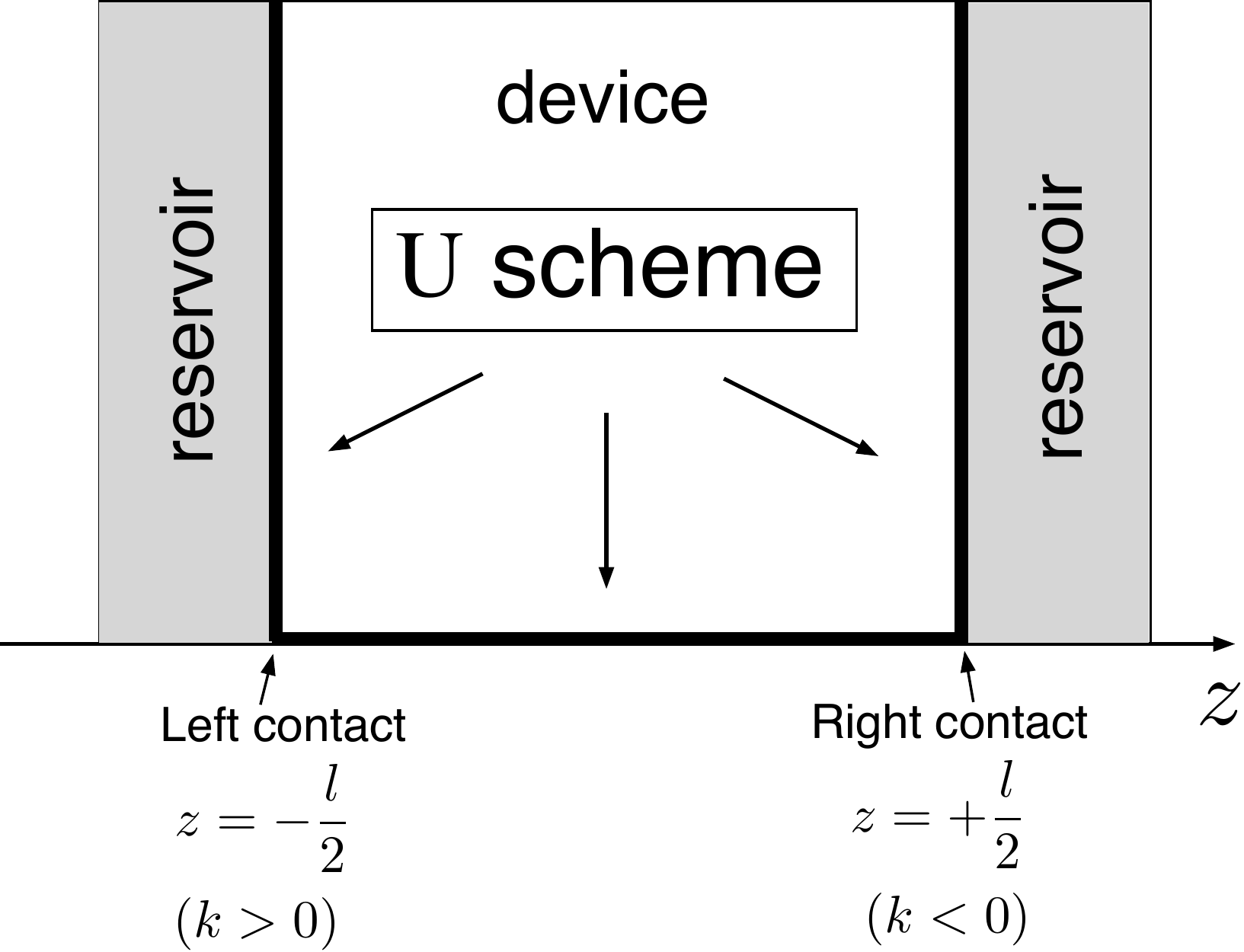}
\caption[]{
The conventional inflow or U boundary condition  scheme adopted in semiclassical  device modeling,\cite{Rossi11} for a one-dimensional problem. The value of the Wigner function $f(z,k)$ is specified at the boundaries $z^b(k)$ of the active region, i.e., $f(-l/2,k > 0)$ and $f(+l/2,k < 0)$ are fixed by the incoming/inflowing carrier distribution function.   
}
\label{fig2}       
\end{figure}

The density-matrix formalism recalled so far --as well as its Weyl-Wigner representation-- applies to ``extended systems'', i.e., systems extending over the whole coordinate space. Indeed, given the state of the system at the initial time $t_0$, its time evolution is fully dictated by the density-matrix equation (\ref{SBE-op}) or, equivalently, via the corresponding Wigner-function equation (\ref{WE2}) defined over the whole coordinate space $\mathbf{r}$.
However, such approach cannot be straightforwardly applied to a nanostructured device, since the latter is a ``localized system'', i.e., a portion of material characterized by a well defined volume and by spatial boundaries acting as electric contacts to external charge reservoirs (see Fig.~\ref{fig1}). 
It follows that, in addition to the initial condition previously mentioned, one is forced to impose on the Wigner-function equation (\ref{WE2}) spatial boundary conditions as well.
Starting from the pioneering work by Frensley,\cite{Frensley86} this has been typically realized by imposing on the Wigner equation the conventional inflow or U boundary condition scheme of the semiclassical device modeling;\cite{Rossi11} the latter amounts to arbitrarily choose/fix the value of an inflowing semiclassical (i.e., positive-definite) carrier distribution $f^b(\mathbf{k})$, regarding the latter as the value of the Wigner function entering the device volume from the spatial boundary $\mathbf{r}^b$, i.e.,
$f^b(\mathbf{k}) \equiv f(\mathbf{r}^b,\mathbf{k})$.
For the particular and relevant case of a one-dimensional problem, schematically depicted in Fig.~\ref{fig2}, this amounts to fixing the value of the incoming Wigner function $f(z,k)$ at the boundaries $z^b(k)$ of the active region, i.e., $f(-l/2,k > 0)$ and $f(+l/2,k < 0)$. 

While such boundary-condition scheme is fully compatible with the conventional semiclassical transport theory (mainly due to the local character of the 
Boltzmann equation), its application to a quantum-mechanical problem is in general not justified.
In particular, two crucial issues need to be investigated:
i) does such classical-like boundary-condition scheme applied to the Wigner-function equation (\ref{WE2})  provide a physically acceptable solution, i.e., a Weyl-Wigner transform of a single-particle density matrix? ii)
is the uniqueness of such   solution  guaranteed?
The aim of this section is to provide a definite answer to the first question, while the second issue will be addressed in Sect.~\ref{s-CL}.

In order to gain more insight about physical versus unphysical solutions, let us focus on the steady-state version of Eq.~(\ref{WE2}), namely
\begin{equation}\label{WE2ss}
\int d\mathbf{r}^\prime \, d\mathbf{k}^\prime 
L(\mathbf{r},\mathbf{k};\mathbf{r}',\mathbf{k}') 
f(\mathbf{r}',\mathbf{k}') = 0
\end{equation}
with
\begin{equation}\label{L}
L(\mathbf{r},\mathbf{k};\mathbf{r}',\mathbf{k}') 
=
\epsilon(\mathbf{r},\mathbf{k};\mathbf{r}',\mathbf{k}') 
+
\Gamma(\mathbf{r},\mathbf{k};\mathbf{r}',\mathbf{k}')\ . 
\end{equation}
Generally speaking, it is a matter of fact that the set of solutions of a given differential equation is usually larger than the physically acceptable ones. 
Indeed, in view of the linear character of Eq.~(\ref{WE2ss}), given two physical solutions $f_a(\mathbf{r},\mathbf{k})$ and $f_b(\mathbf{r},\mathbf{k})$, the linear combination
\begin{equation}\label{alphabeta}
f(\mathbf{r},\mathbf{k}) = c_a f_a(\mathbf{r},\mathbf{k}) + c_b f_b(\mathbf{r},\mathbf{k})
\end{equation}
is also a mathematical solution of the same equation, leading to a spatial carrier density [see Eq.~(\ref{dens-from-WF})]   of the form 
\begin{equation}\label{nzalphabeta}
n(\mathbf{r}) = c_a \, n_a(\mathbf{r}) + c_b \, n_b(\mathbf{r})\ .
\end{equation}
In spite of the positive-definite character of the spatial charge densities $n_a$ and $n_b$,  an inappropriate choice of the   coefficients $c_a$ and $c_b$ may give rise to a partially negative charge distribution, which corresponds to an unphysical solution. On the other hand,   the presence of given spatial boundary conditions is expected to impose additional constraints on the two coefficients $c_a$ and $c_b$, thus reducing the set of available solutions.

In order to better understand the link among the system density matrix $\hat\rho$, the Wigner function $f(\mathbf{r},\mathbf{k})$, and the corresponding boundary function $f^b(\mathbf{k})$, let us examine in more detail the Weyl-Wigner transform in (\ref{WF-op}). 
Because the density-matrix operator $\hat\rho$ is always hermitian and positive-definite, its spectral decomposition  
\begin{equation}
\hat\rho=\sum_\beta p_\beta \, |\beta\rangle \langle \beta| \label{rho-sp-dec}
\end{equation} 
involves {\it non-negative}  eigenvalues $p_\beta \ge 0$. 
Inserting Eq.~(\ref{rho-sp-dec}) into (\ref{WF-op}) and employing the pure-state result (\ref{WFps}), the Wigner function turns out to be
\begin{equation}\label{WFzbeta}
f(\mathbf{r},\mathbf{k}) = \sum_\beta \, p_\beta \, f_\beta(\mathbf{r},\mathbf{k})\ , \hspace{1cm} p_\beta \ge 0 \quad.
\end{equation}
The above linear combination can be regarded as a statistical average (i.e., a mixed state) of the Wigner functions $f_\beta(\mathbf{r},\mathbf{k})$ corresponding to the pure states $|\beta\rangle$. 
As a consequence the spatial carrier density corresponding to the Wigner function in (\ref{WFzbeta}) reads
\begin{equation}\label{nzbeta}
n(\mathbf{r}) = \sum_\beta p_\beta \, n_\beta(\mathbf{r})\ 
\end{equation}
and is always positive, being a linear combination of the positive-definite functions $n_\beta(\mathbf{r})$ with positive-definite coefficients $p_\beta$. Such physical result thus originates from the positive-definite character of the density-matrix operator.\\

On the other hand, in the conventional boundary-condition scheme (see Fig.~\ref{fig2}) employed for the simulation of quantum devices with open spatial boundaries, one arbitrarily fixes the value of the Wigner function entering the device from the spatial boundary $\mathbf{r}^b$.
The crucial question is whether, for any given (real and positive-definite) boundary function 
$f^b(\mathbf{k}) \equiv f(\mathbf{r}^b,\mathbf{k})$, any mathematical solution of Eq.~(\ref{WE2ss}) is also a physically acceptable one. 
Let us consider a generic solution $\tilde{f}(\mathbf{r},\mathbf{k})$ of the Wigner equation. 
Exploiting  the completeness relation of the  Wigner operators, namely 
\begin{equation}
\label{WWCS}
(2\pi)^{-3}\, {\rm tr} \left\{ \hat{W}^{}(\mathbf{r},\mathbf{k}) \, \hat{W}^\dagger(\mathbf{r}^\prime,\mathbf{k}^\prime) \right\} = \delta(\mathbf{r}-\mathbf{r}^\prime) \,\delta(\mathbf{k}-\mathbf{k}^\prime)\ ,
\end{equation}
the (real) function $\tilde{f}(\mathbf{r},\mathbf{k})$ defined on the phase space can always be written as
\begin{equation}
\label{hatF}
\tilde{f}(\mathbf{r},\mathbf{k})= {\rm tr}\left\{ \hat{W}(\mathbf{r},\mathbf{k}) \, \hat{\Phi}\right\} \quad.
\end{equation}
Here $\hat{\Phi}$ is an {\it hermitian} operator, defined as   
\begin{equation}
\hat{\Phi} = (2\pi)^{-3}\, \int\!\!\!   \int d\mathbf{r} \, d\mathbf{k} \, \hat{W}(\mathbf{r},\mathbf{k}) \,  \tilde{f}(\mathbf{r},\mathbf{k}) \quad,
\end{equation}
whose spectral decomposition
\begin{equation}
\hat{\Phi}=\sum_{\tilde{\beta}} e_{\tilde{\beta}} \, |\tilde{\beta} \rangle \langle \tilde{\beta}|
\end{equation}
involves pure states $|\tilde{\beta} \rangle$ and, due to hermiticity, real eigenvalues $e_{\tilde{\beta}}$.
However, in contrast to the case of a density-matrix operator $\hat\rho$ [see Eq.(\ref{rho-sp-dec})], the eigenvalues of $\hat{\Phi}$ are not necessarily positive. 
Denoting by $f_{\tilde{\beta}}(\mathbf{r},\mathbf{k})$  the Wigner function of the pure state $|\tilde{\beta}\rangle$, the  solution~(\ref{hatF}) can be written as
\begin{equation}\label{WFzbetabar}
\tilde{f}(\mathbf{r},\mathbf{k}) = \sum_{\tilde{\beta}} \, e_{\tilde{\beta}} \, f_{\tilde{\beta}}(\mathbf{r},\mathbf{k})\ , 
\end{equation}
which is  not necessarily a mixed-state Wigner function of the form (\ref{WFzbeta}), due to the possible presence of negative eigenvalues $e_{\tilde{\beta}}$. It may therefore  be unphysical, since the corresponding spatial carrier density 
\begin{equation}\label{nzbetabar}
\tilde{n}(\mathbf{r}) = \sum_{\tilde{\beta}} \, e_{\tilde{\beta}} \, n_{\tilde{\beta}}(\mathbf{r})
\end{equation}
is not necessarily positive-definite.

This is the mathematical explanation of the   unphysical results reported in Ref.~[\onlinecite{Taj06}] as well as in Fig.~\ref{fig8}.
As we shall discuss in Sects.~\ref{s-CL} and \ref{s-SC}, from a physical point of view the presence of such unphysical solutions is a clear indication that the non-local character of the Liouville superoperator in (\ref{L}) does not allow one to arbitrarily choose/fix the boundary values of the unknown Wigner function regardless of the specific device under examination, since the Wigner function in $\mathbf{r}$ depends, in general, on the value of the device potential profile in any other point $\mathbf{r}'$.

To summarize, in view of the completeness property in (\ref{WWCS}), it is always possible to identify a proper linear combination (\ref{WFzbetabar}) of the pure-state Wigner functions fulfilling the desired boundary values $f^b(\mathbf{k})$. However, such linear combination does not necessarily correspond to a physically acceptable solution (see also Sec.~\ref{s-IERDP}).
Moreover, while the existence of such mathematical solution is guaranteed, its uniqueness strongly depends  on the particular properties of the effective Liouville superoperator $L$ in (\ref{L}); in particular, as discussed in the following section, in the   coherent limit the solution of Eq.~(\ref{WE2ss}) (compatible with given spatial boundaries) is not unique.

\section{The coherent limit}\label{s-CL}

In order to  investigate the intrinsic limitations of the boundary-condition scheme pointed out above, we shall first focus on a fully coherent system/device, where energy-dissipation/decoherence processes occur over timescales that are much longer than the typical timescales induced by $\hat{H}^\circ$. In this regime, the density-matrix equation (\ref{SBE-op}) reduces to the Liouville-von Neumann equation
\begin{equation}\label{DMECL}
{d \hat\rho \over dt} = 
{1 \over i\hbar}\,\left[\hat H^\circ, \hat\rho\right]\ .
\end{equation}

\subsection{The Wigner transport equation}\label{ss-WE}

For the purpose of the present article, it is enough to consider a one-dimensional system ($\mathbf{r},\mathbf{k} \to z,k$) described by the envelope-function Hamiltonian\cite{Rossi11}
 \begin{equation}\label{EFH}
\hat H^\circ = K(\hat{k}) + V\left(\hat z\right)\ ,
\end{equation}
where $\hat z$ and $\hat k$ denote, respectively, the quantum-mechanical operators associated to the electronic coordinate ($z$) and to the electronic momentum/wavevector ($k$); generalizations to a fully three-dimensional problem are straightforward. According to the usual prescription of the envelope-function theory, the function $K$ in Eq.~(\ref{EFH}) describes the bulk electronic band, while $V$ describes the nanostructure  potential profile.  
The  Hamiltonian (\ref{EFH}) leads Eq.~(\ref{DMECL}) to acquire the form 
\begin{equation}\label{DMECLbis}
{d \hat\rho \over dt} =
\left. {d \hat\rho \over dt}\right|_K
+ 
\left. {d \hat\rho \over dt}\right|_V \, ,
\end{equation}
with
\begin{equation}\label{DMEK}
\left. {d \hat\rho \over dt}\right|_K = 
{1 \over i\hbar}\,\biggl[K(\hat{k}), \hat\rho\biggr]
\end{equation}
and
\begin{equation}\label{DMEV}
\left. {d \hat\rho \over dt}\right|_V = 
{1 \over i\hbar}\,\biggl[V(\hat z ), \hat\rho\biggr]\ .
\end{equation}
Applying the Weyl-Wigner transform to the density-matrix equation~(\ref{DMECLbis}),  one gets the Wigner-function equation for $f(z,k)$. In doing that,  
 the Wigner function~(\ref{WF-op}) can be expressed in two different and equivalent ways, corresponding to the momentum 
($k$) and coordinate ($z$) representations, respectively. 
By setting $\alpha=k$ as well as $\alpha=z$ in Eq.~(\ref{WF}) one obtains
\begin{eqnarray} 
f(z,k) &=& \int dk' e^{i zk'} \, \rho\left(k+{k' \over 2},k-{k' \over 2}\right) \ \label{WWk} \\
  &=& \int dz' e^{-i k z'}\, \rho\left(z+{z' \over 2},z-{z' \over 2}\right) \ .
\label{WWz}
\end{eqnarray}
These two expressions turn out to be both useful because the kinetic and potential contributions (\ref{DMEK}) and (\ref{DMEV}) are diagonal in the momentum  ($k$) and coordinate ($z$) representations, respectively, i.e.
\begin{equation}\label{DMEk}
\left. {d \rho(k_1,k_2) \over dt}\right|_K = 
{K(k_1)-K(k_2) \over i\hbar}\,\rho(k_1,k_2)
\end{equation}
and
\begin{equation}\label{DMEz}
\left. {d \rho(z_1,z_2) \over dt}\right|_V = 
{V(z_1)-V(z_2) \over i\hbar}\,\rho(z_1,z_2)\ .
\end{equation}

i) By applying the Weyl-Wigner transform (\ref{WWk}), as well as its inverse (given by Eq.~(\ref{WF-op-inv}) written in the $k$-representation), 
\begin{equation}\label{IWWk}
\rho\left(k+{k' \over 2},k-{k' \over 2}\right) = 
\int dz {e^{-i k' z} \over 2\pi}\,f(z,k)\ ,
\end{equation}
to the kinetic contribution   (\ref{DMEk}), one gets
\begin{equation}\label{WEk}
\left. {\partial f(z,k) \over \partial t}\right|_K
= 
-\,\int dz'\,{\cal K}(z-z',k) f(z',k)
\end{equation}
with
\begin{equation}\label{calK}
{\cal K}(z'',k)\! =\! {i \over \hbar} \int \!dk'
\frac{e^{i z'' k'}}{2\pi}
\left[\!K\left(k+{k' \over 2}\right)\!-\!K\left(k-{k' \over 2}\right)\!\right] \ .
\end{equation}
The  kinetic operator  in the Wigner picture, appearing on the r.h.s. of Eq.~(\ref{WEk}), is always local in $k$ and, in general, is non-local in $z$. In particular, by adopting the usual effective-mass approximation,  
\begin{equation}\label{EMA}
K(k) = {\hbar^2k^2 \over 2m^*}\ ,
\end{equation}
the non-local kinetic operator (\ref{WEk}) reduces to
\begin{equation}\label{WEkEMA}
\left. 
{\partial f(z,k) \over \partial t}\right|_K
= 
-v(k)\,{\partial f(z,k) \over \partial z}\ ,
\end{equation}
where $v(k) = {\hbar k/m^*}$ denotes the effective-mass carrier group velocity.
Notably, within the effective-mass approximation (\ref{EMA}) the kinetic contribution coincides with its semiclassical counterpart, i.e., it reduces to the usual diffusion term of the Boltzmann equation.\\

ii) By applying the Weyl-Wigner transform (\ref{WWz}), as well as its inverse (given by Eq.~(\ref{WF-op-inv}) written in the $z$-representation), 
\begin{equation}\label{IWWz}
\rho\left(z+{z' \over 2},z-{z' \over 2}\right) = 
\int dk {e^{i z' k} \over 2\pi}\,f(z,k)\ ,
\end{equation}
to the potential contribution in (\ref{DMEk}), one gets
\begin{equation}\label{WEz}
\left. {\partial f(z,k) \over \partial t}\right|_V
= 
-\,\int dk'\,{\cal V}(z,k-k') f(z,k')
\end{equation}
with
\begin{equation}\label{calV}
{\cal V}(z,k'')  =  {i \over \hbar} \int\! dz'
\frac{e^{-i k'' z'}}{2\pi}
\left[\!V\left(z+{z' \over 2}\right)\!-\!V\left(z-{z' \over 2}\right)\!\right] \ .
\end{equation}
Oppositely to the kinetic one, the potential operator appearing on the r.h.s. of Eq.(\ref{WEz}) is always local in $z$ and, in general, is non-local in $k$.
For the particular case of a quadratic potential
\begin{equation}\label{QPP}
V(z) = {1 \over 2} a z^2 + b z + c \quad,
\end{equation}
corresponding to the classical force
\begin{equation}\label{lf}
F(z) = -{d V(z) \over dz} = -(a z + b) ,
\end{equation}
the non-local potential operator (\ref{WEz}) simply reduces to 
\begin{equation}\label{WEkPP}
\left. {\partial f(z,k) \over \partial t}\right|_V
= 
-\,{F(z) \over \hbar}\,{\partial f(z,k) \over \partial k}\ .
\end{equation}
Thus, for the particular case of the quadratic potential profile (\ref{QPP}), the potential contribution coincides with its semiclassical counterpart, i.e., it reduces to the standard drift term of the Boltzmann equation; it follows that the non-local character of the generic potential superoperator in (\ref{WEz}) vanishes in the presence of a parabolic potential only.\\

The analysis performed so far has  shown a strongly symmetric role between real-space ($z$) and momentum ($k$) coordinates; this is confirmed by the fact that the corresponding equations of motion (each one written within the related representation) display the very same mathematical structure [see Eqs.~(\ref{DMEk}) and (\ref{DMEz})].
Moreover, for a physical system characterized by an effective Hamiltonian quadratic in both  the coordinate and the momentum, the equation of motion of the Wigner function coincides with its semiclassical (Boltzmann) counterpart, thus showing the intimate link between the Wigner function and the semiclassical distribution.
This can also be regarded as a formal proof of the fact that, for a  particle subjected to a quadratic potential, its classical and quantum equations of motion coincide, a fundamental result originally pointed out by Richard P. Feynman via his ``path integral'' formulation of quantum mechanics.\cite{Feynman65} \\

For the microscopic modeling of semiconductor quantum devices, the effective-mass approximation   (\ref{EMA}) is widely employed, and constitutes a good starting point for the description of the bulk band structure. In contrast, for a generic optoelectronic device, the effective potential profile $V(z)$ is usually far from the quadratic form in (\ref{QPP}).
As a consequence, within this approximation scheme the   single-particle superoperator $\epsilon$ in (\ref{WE2epsilon}) is always local in $z$, and the total (i.e., kinetic plus potential) equation of motion for $f(z,k)$ --obtained combining Eqs.~(\ref{WEkEMA}) and (\ref{WEz})-- contains a non-local term in $k$ induced by the potential profile:
\begin{equation}\label{WECL}
{\partial f(z,k) \over \partial t}
+
v(k)\,{\partial f(z,k) \over \partial z} + \int dk' {\cal
V}(z,k-k') f(z,k') = 0\ .
\end{equation}

Equation (\ref{WECL}), also referred to as the Wigner transport equation, describes the time evolution of the one-dimensional Wigner function in the absence of energy-dissipation/decoherence processes. In steady-state conditions ($\partial f(z,k) / \partial t  = 0$) it reduces to  
\begin{equation}\label{WECLSS}
v(k)\,{\partial f(z,k) \over \partial z} = - \int dk' {\cal
V}(z,k-k') f(z,k')\ .
\end{equation}
In terms of the variable $z$ the above equation is a first-order differential equation. In this respect, it is thus similar to the (steady-state) semiclassical Boltzmann equation\cite{Rossi11}
\begin{equation}
\label{BE}
v(k)\,{\partial f(z,k) \over \partial z} = -{F(z) \over \hbar} \, \frac{\partial f(z,k)}{\partial k}\quad.
\end{equation}
Based on this analogy,  outlined in the pioneering work by Frensley,\cite{Frensley86} several   quantum-transport problems have been treated by following a semiclassical approach, i.e. by applying to the Wigner transport equation (\ref{WECLSS}) the   strategy commonly adopted for the Boltzmann Equation~(\ref{BE}). Indeed most of these studies\cite{Frensley90} are based on a numerical solution of Eq.~(\ref{WECLSS}), often supplemented by an additional relaxation-time term (see Sec.~\ref{s-IERDP}), where one imposes on $f(z,k)$ the U spatial boundary condition scheme described in Sec.~\ref{s-U}. 
The latter,   depicted in Fig.~\ref{fig2}, consists in requiring that the inflowing Wigner function acquires some fixed values  at the two contacts $z=\pm l/2$, and that these values are determined by the distribution of carriers incoming from the two reservoirs. Explicitly, the values  $f(-l/2,k)$ are specified  for carriers incoming from the left reservoir ($k>0$) and the values $f(+l/2,k)$ are specified for electrons incoming from the right reservoir ($k<0$). In a compact notation, introducing $z^b(k)=-\mbox{\rm sign}(k) \, l/2$, the Wigner transport equation (\ref{WECLSS}) is thus equipped with the $k$-dependent spatial boundary condition 
\begin{equation}
\label{bc}
f^b(k) \equiv f(z^b(k),k)\ .
\end{equation} 

Within such boundary-condition paradigm, it is also possible to rewrite the Wigner problem (\ref{WECLSS})-(\ref{bc}) in an equivalent integral form\cite{Taj06}
\begin{eqnarray}
\displaystyle v(k) \, f(z,k)  &= &\displaystyle v(k) \, f^b(k) -\label{WE-int} \\
& & -\int_{z^b(k)}^z dz^\prime  \int_{-\infty}^{+\infty} dk^\prime \, \mathcal{V}(z^\prime,k-k^\prime) \, f(z,k^\prime) \, .\nonumber 
\end{eqnarray}
This integral equation is the starting point of the Neumann-series solution employed in Ref.~[\onlinecite{Taj06}], i.e., a numerical treatment based on an iterative expansion of the solution $f(z,k)$ in powers of the potential superoperator~$\mathcal{V}$.

We wish to point out that, in spite of the (classical versus quantum) analogies mentioned above, an important difference emerges between the Wigner equation (\ref{WECLSS}) and the semiclassical Boltzmann equation (\ref{BE}). While   the latter is local in $k$, the former is not.  Indeed, because of the non-local character (in $k$) of the potential superoperator $\mathcal{V}$ appearing in (\ref{WECLSS}), the differential equation for one value of $k$ is in fact coupled to the differential equations for all other $k$ values. The non-locality of $\mathcal{V}$ therefore makes the Wigner problem intrinsically different from the Boltzmann one.
Indeed, while the solution of the Boltzmann equation (compatible with given boundary values) is always unique, the same does not apply to the Wigner equation (see below).

\subsection{Non-uniqueness of the solution}\label{ss-NU}

In order to show the non-uniqueness of the solution of the Wigner transport equation (\ref{WECLSS}), we start by investigating its general symmetry properties.
A closer inspection of the Wigner potential in (\ref{calV}) reveals its antisymmetric nature with respect to the momentum coordinate, i.e.,
\begin{equation}\label{calVask}
{\cal V}(z,k'') = -\,{\cal V}(z,-k'')\ .
\end{equation}
As a consequence, both $f(z,k)$ and $f(z,-k)$ are solutions of the Wigner equation (\ref{WECLSS}).
Such property reflects the time-reversal symmetry, i.e., it corresponds to the fact that for any given solution $\phi(z)$ of the time-dependent Schr\"odinger equation, its complex conjugate $\phi^*(z)$ is also a solution; this is confirmed by recalling that, for a pure state $|\phi\rangle \langle\phi|$ corresponding to a wavefunction $\phi(z)$, the related Wigner function (\ref{WF-op}) is simply given by
\begin{equation}\label{WFphi}
f(z,k) = \int dz'\, 
\phi^{ }\left(z+{z' \over 2}\right)
\,e^{-i k z'}\, 
\phi^*\left(z-{z' \over 2}\right) \ ,
\end{equation}
and noticing that the replacement in (\ref{WFphi}) of the two wavefunctions with their complex conjugates is  equivalent to changing $k$ in $-k$.

In addition to the antisymmetry (\ref{calVask}) w.r.t. $k$, in the presence of a spatially symmetric potential $V(z) = V(-z)$, the Wigner potential (\ref{calV}) turns out to be antisymmetric with respect to the spatial coordinate as well,  
\begin{equation}\label{calVasz}
{\cal V}(z,k'') = -\,{\cal V}(-z,k'')\ ,
\end{equation}
implying that, for a given solution $f(z,k)$ of the Wigner equation (\ref{WECLSS}), also $f(-z,k)$ is a solution of the same equation.
Such property corresponds to the fact that, in the presence of a symmetric potential, for any given solution $\phi(z)$, the wavefunction $\phi^*(-z)$ is a solution as well.

For any finite and piece-wise-constant potential $V(z)$, one can easily define a set of doubly degenerate eigenstates called scattering states.\cite{Datta97} For the sake of simplicity, let us assume that $V(z \to -\infty) = V(z \to +\infty) = 0$; in this case, for any positive energy value $\epsilon$ it is possible to define two degenerate eigenstates, 
usually referred to as left and right scattering states, corresponding, respectively, to a  plane wave incoming from  left ($\overline{k}>0$) and right ($\overline{k}<0$), with unit  amplitude and wavevector 
\begin{equation}\label{kk}
\kk = \pm {\sqrt{2 m^* \epsilon} \over \hbar}
\end{equation}
(a typical example will be discussed in Sec.~\ref{ss-db}).
We shall thus label this specific set of eigenfunctions of the effective Hamiltonian (\ref{EFH}) via the continuous quantum number $\kk$ as ${\phi}_\kk(z)$.
Recalling the pure-state prescription in (\ref{WFphi}), the Wigner function corresponding to the generic scattering state 
is given by
\begin{equation}\label{WFkk}
{f}_\kk(z,k) = 
\int dz' 
{\phi}_\kk\left(z\!+\!{z' \over 2}\right) 
e^{-i kz'} 
{\phi}^*_\kk\left(z\!-\!{z' \over 2} \right) \ .
\end{equation}

Taking into account that for any value $\kk$ the function ${f}_\kk(z,k)$ is a solution of the Wigner equation (\ref{WECLSS}), and that the latter is linear and homogeneous, it follows that any function
\begin{equation}\label{sol1}
f(z,k) = \int d\kk \, a(\kk) \,  {f}_\kk(z,k)
\end{equation}
is itself a solution.

In order to verify if  the function in (\ref{sol1}) is a unique solution of the Wigner equation (\ref{WECLSS}) compatible with the given spatial boundary condition in (\ref{bc}), we impose that the generic solution (\ref{sol1}) on the spatial boundary $z = z^b(k)$   assumes the required boundary value $f^b(k)$, i.e.,
\begin{equation}\label{fbbis}
f\left(z^b(k),k\right) = \int d\kk \, a(\kk) \,  {f}_\kk\left(z^b(k),k\right) = f^b(k)\ .
\end{equation}
This can be regarded to as an infinite set of linear equations for the infinite set of unknowns $a(\kk)$:
\begin{equation}\label{ls1}
\int d\kk \, L^a(k,\kk) \, a(\kk) = f^b(k)
\end{equation}
with
\begin{equation}\label{La}
L^a(k,\kk) = {f}_\kk\left(z^b(k),k\right)\ .
\end{equation}
Assuming that the linear operator $L^a$ is non-singular, there is always a unique choice of the coefficients $a(\kk)$ compatible with the desired boundary conditions (\ref{bc}), and therefore a unique solution $f(z,k)$ of the Wigner equation (\ref{WECLSS}).

Importantly, the above conclusion is based on the assumption that the function $f(z,k)$ in (\ref{sol1}) is the most general solution of the Wigner equation; in what follows we shall show that this assumption is   wrong. 
Indeed in view of the time-reversal symmetry $z,k \to z, -k$ mentioned above, in addition to the set of eigenvalue Wigner functions ${f}_\kk(z,k)$ in (\ref{sol1}), one may consider a second (and linearly independent) set of solutions given by ${f}_\kk(z,-k)$.
This allows one to extend the set of possible solutions in (\ref{sol1}) as
\begin{equation}\label{sol2}
f(z,k) = \int d\kk \left(
a(\kk) {f}_\kk(z,k)
+
b(\kk) {f}_{-\kk}(z,-k)
\right)\ ,
\end{equation}
whose spatial charge density is given by
\begin{equation}\label{nz}
n(z) = \int d\kk \left(
a(\kk)  {n}_\kk(z)
+
b(\kk)  {n}_{-\kk}(z)
\right)
\end{equation}
with ${n}_\kk(z) = \left| {\phi}_\kk(z)\right|^2$.

By imposing once again the boundary-value prescription (\ref{bc}) on the new generic solution in (\ref{sol2}) one gets
\begin{equation}\label{ls2}
\int d\kk \left(
L^a(k,\kk) a(\kk) 
+
L^b(k,\kk) b(\kk)
\right) 
= f^b(k)
\end{equation}
with
\begin{equation}\label{Lb}
L^b(k,\kk) = {f}_{-\kk}\left(z^b(k),-k \right)\ .
\end{equation}
In Eq.(\ref{ls2})  a second (infinite) set of unknown quantities $b(\kk)$ appears, in addition to the (infinite) set of unknown quantities $a(\kk)$. It follows that, differently from the linear problem in (\ref{ls1}), 
the new global set of coefficients 
$\{a(\kk), b(\kk)\}$ 
is not uniquely determined by the corresponding linear set of equations in (\ref{ls2}).\cite{note-degeneracy}

Among all possible choices of the coefficients, it is useful to consider the particular class of solutions 
\begin{equation}\label{c}
b(\kk) = c \, a(\kk)
\end{equation}
parameterized by the real number $c$.
In this case, the generic solution in (\ref{sol2}) reduces to
\begin{equation}\label{sol3}
f(z,k) = \int d\kk \, a(\kk) \, {g}_\kk(z,k)
\end{equation}
with 
\begin{equation}\label{ffs}
g_\kk(z,k) = 
{f}_\kk(z,k) + c \, {f}_{-\kk}(z,-k)\ ,
\end{equation}
and the corresponding linear problem in (\ref{ls2}) reduces to
\begin{equation}\label{ls3}
\int d\kk \,
L^c(k,\kk) \, a(\kk) 
= f^b(k)
\end{equation}
with
\begin{equation}\label{Lc}
L^c(k,\kk) 
=
L^a(k,\kk) + c \,L^b(k,\kk) \ ,
\end{equation}
thus providing --for any given value of the parameter $c$-- a unique value of the coefficients $a(\kk)$ compatible with the desired boundary conditions.

Let us finally discuss the non-uniqueness of the solution in the presence of a spatially symmetric potential: $V(z) = V(-z)$. Indeed in this case it is possible to show that, changing $z$ into $-z$, left-scattering states map into right-scattering ones, and vice versa, i.e., 
${\phi}_\kk(z) =  {\phi}_{-\kk}(-z)$.
In terms of the Wigner-function picture $z,k$, this symmetry property reduces to
\begin{equation}\label{sym2}
{f}_{-\kk}(z,-k) = {f}_\kk(-z,k)\ .
\end{equation}
Employing this symmetry property, the generic solution (\ref{sol2}) turns out to be
\begin{equation}\label{sol2bis}
f(z,k) = \int d\kk \left(
a(\kk) {f}_\kk(z,k)
+
b(\kk) {f}_\kk(-z,k)
\right)\ ,
\end{equation}
and the spatial charge distribution in (\ref{nz}) reduces to
\begin{equation}\label{nzbis}
n(z) = \int d\kk \left(
a(\kk)  {n}_\kk(z)
+
b(\kk)  {n}_\kk(-z)
\right)\ .
\end{equation}
Also in the presence of a spatially symmetric potential, the above generic solution compatible with given boundary conditions is definitely not unique; it is then useful to consider the solution set in (\ref{c}) for $c = 1$, i.e., $b(\kk) = a(\kk)$; in this particular case, combining the definition in (\ref{ffs}) with the symmetry property in (\ref{sym2}), one gets
\begin{equation}\label{ffsbis}
{g}_\kk(z,k) = 
{f}_\kk(z,k) + {f}_\kk(-z,k)\ ,
\end{equation}
i.e., the functions $g$ entering the linear combination (\ref{sol3}) in this case are always spatially symmetric, and are simply given by twice the symmetric part of the scattering state Wigner function ${f}_\kk$; 
it follows that for the particular case/choice $c = 1$ the generic solution in (\ref{sol3}) is always spatially symmetric, and the same applies to the corresponding charge density in (\ref{nzbis}).

Recalling that  in the presence of a spatially symmetric potential the analytical and  numerical results reported in Ref.~[\onlinecite{Taj06}] (based both on a symmetric finite-difference solution of Eq.~(\ref{WECLSS}) and on a Neumann-series expansion of Eq.~(\ref{WE-int}) (see below)) 
correspond to spatially symmetric Wigner functions only ($f(z,k) = f(-z,k)$), 
the natural conclusion is that, among the infinite set of coefficients compatible with the given boundary conditions, such treatments automatically provide/select the symmetric choice $c = 1 \to b(\kk) = a(\kk)$.
Moreover, since these treatments show a continuous transition of the solution moving from a symmetric to a non-symmetric potential, one is forced to conclude that also in the presence of non-symmetric potentials the numerical approaches just mentioned are expected to provide/select again the particular solution $b(\kk) = a(\kk)$ in (\ref{sol2}), which in general is spatially non-symmetric.
The existence of an infinite set of degenerate solutions, i.e., solutions compatible with the same boundary values, allows also to explain the significant discrepancies between finite-difference treatments based on different (spatially symmetric versus non-symmetric) discretization schemes, already pointed out in Ref.~[\onlinecite{Taj06}]: 
a change in the spatial discretization scheme may induce significant changes in the numerical results, since, regardless of the actual grid size, it may select a new solution (i.e., a different value of the parameter $c$ in Eq.~(\ref{c})) within the degenerate subspace.

Let us finally discuss the non-uniqueness of the solution in terms of the integral version of the Wigner equation (\ref{WE-int}). To this end, by adopting a compact notation, the latter can be written as
\begin{equation}\label{e:FS2}
\mathcal{A} f = f^b
\end{equation}
with 
\begin{eqnarray}
\mathcal{A} \, f(z,k) &\doteq &   f(z,k) + \\ & & +\int_{z^b(k)}^{z} dz^\prime  \int_{-\infty}^{+\infty} dk^\prime \, \frac{\mathcal{V}(z^\prime,k-k^\prime)}{v(k)} \,  f(z^\prime,k^\prime) \, .\nonumber
\end{eqnarray}
The non-uniqueness of the solution previously shown tells us that the Wigner superoperator $\mathcal{A}$ is necessarily not invertible, which implies that the well-known Neumann series/expansion
\begin{equation}\label{NE}
f = \mathcal{A}^{-1} f^b
= \sum_{n=0}^\infty \left(1-\mathcal{A}\right)^n f^b 
\end{equation}
in this case provides/selects just one of the infinite solutions; in particular, as shown in Ref.~[\onlinecite{Taj06}], for a symmetric potential the result of the above Neumann expansion is always symmetric, which corresponds to the particular choice 
$b(\kk) = a(\kk)$ previously discussed.

\subsection{Example: The case of a delta-like potential}\label{ss-db}

As an analytically solvable model, let us consider the case of the delta-like potential barrier 
\begin{equation}\label{delta}
V(z) = \Lambda \, \delta(z) \quad, 
\end{equation}
where $\Lambda$ denotes the barrier-strength parameter. 
Within the effective-mass approximation (see Eq.~(\ref{EMA})) the delta-barrier Schr\"odinger equation (corresponding to the envelope-function Hamiltonian (\ref{EFH})) reads
\begin{equation}\label{Schr}
\left[-{\hbar^2 \over 2m^*}\,{\partial^2 \over \partial z^2} + \Lambda \delta(z) \right] \phi(z) = \epsilon \, \phi(z)\ .
\end{equation}
As discussed above, the latter exhibits a continuous set of doubly-degenerate scattering eigenstates ${\phi}_\kk(z)$ parameterized by the continuous quantum number $\kk$ in (\ref{kk}) and describing (for any given energy $\epsilon = {\hbar^2\kk^2 / 2m^*}$) 
injection onto the barrier from the left side ($\overline{k}>0 \,$, left-scattering state) and from the right side ($\overline{k}<0 \,$, right-scattering state).
The explicit form of these scattering states corresponding to the delta-like potential (\ref{delta}) can be written in a compact way as
\begin{equation}\label{ss+-}
{\phi}_\kk(z) =
{1 \over \sqrt{\Omega}}\,
\cases{
 e^{i \kk z} + r_\kk \, e^{-i \kk z}
 & {\rm for} \quad $\kk z < 0$ \cr
  t_\kk \, e^{i \kk z}
 & {\rm for} \quad $\kk z > 0$
}\ ,
\end{equation}
Here
\begin{equation}\label{RT}
r_\kk = - {i\lambda_\kk \over 1+ i\lambda_\kk}\ , \qquad t_\kk = {1 \over 1+i\lambda_\kk}  
\end{equation}
denote the reflection and transmission amplitudes, respectively,  
\begin{equation}\label{lambda}
\lambda_\kk = {m^* \Lambda  \over \hbar^2 |\kk|}
\end{equation}
is a dimensionless barrier-strength parameter, and   the prefactor $1/\sqrt{\Omega}$ ensures the normalization of the above scattering states over the whole system (device+reservoirs) length $\Omega$.\cite{note-normalizz}

Let us now focus on the pure-state Wigner function corresponding to the generic  left-scattering state ($\kk > 0$) in (\ref{ss+-}). 
By inserting its explicit form 
into the general prescription (\ref{WFkk}), a lengthy but straightforward calculation, outlined in the Appendix, leads to obtain 
\begin{widetext}
\begin{eqnarray}
{f}_{\kk>0}(z,k) & = &
{2\pi \over \Omega}\, 
\left[
T_\kk \, \delta(k-\kk)
+ i t_\kk r^*_\kk  \left( \sin(2 \kk z) \delta(k)
-  
{2 \kk \cos\left(2(\kk-k) z\right) 
\over 
2\pi k (\kk-k)}
\right) \right. \nonumber \\
& & \hspace{1cm}
\left. 
- {\theta(-z) R_\kk \over \pi}
\left(
{\sin\left(2(\kk-k)z\right) \over \kk-k}
+
{\sin\left(2(\kk+k) z\right) \over \kk+k}
-
{2 \cos(2 \kk z) \sin(2 k z) \over k}
\right)
\right]\ ,\label{WFss2}
\end{eqnarray}
\end{widetext} 
where $R_\kk = \left|r_\kk\right|^2$ and $T_\kk = \left|t_\kk\right|^2$ are the usual reflection and transmission coefficients.

It is worth noticing that the scattering-state Wigner function (\ref{WFss2}) is spatially non-symmetric, similarly to the charge density 
(${n}_\kk(z) = \left|{\phi}_\kk(z)\right|^2$)
corresponding to the generic scattering-state wavefunction in (\ref{ss+-}):
\begin{equation}\label{nzss2}
{n}_\kk(z) \!=\!
n_0
\cases{
1\!+\!R_\kk\!+ 2  \Re  \left[ r_\kk  e^{-2 i \kk z} \right]
& {\rm for}  $\kk z < 0$ \cr
T_\kk
& {\rm for}  $\kk z > 0$
}.
\end{equation}
This asymmetry has a physically intuitive explanation: since a left/right scattering state describes carrier injection from left/right, the presence of the barrier causes a charge accumulation on the left/right of the barrier with respect to the density of the carriers transmitted on the right/left. Here, as well as throughout the whole article, $n_\circ$ denotes the (space-independent) charge density corresponding to the barrier-free case. For the case we are presently considering --that is, one single scattering state as in Eq.~(\ref{ss+-})-- $n_0$ is simply given by $1/\Omega$.\\

A lengthy but straightforward calculation, summarized in the Appendix, allows one to verify that the Wigner function (\ref{WFss2})   is a solution of the  Wigner transport equation (\ref{WECLSS}), where   the   Wigner potential $\mathcal{V}(z,k)$ induced by the delta-like barrier (\ref{delta})  is now given by
\begin{equation}\label{calVdelta}
{\cal V}(z,k) = - {4\Lambda  \over 2 \pi \hbar} \sin\left(2 k z\right) \ ,
\end{equation}
as can be easily verified via its definition in Eq.~(\ref{calV}).
As shown in Sec.~\ref{ss-NU}, the analytical solution in (\ref{WFss2}) --compatible with its boundary values $f^b(k) = {f}_{\kk>0}(z^b(k),k)$-- is definitely not unique. Indeed, adopting once again the compact notation introduced in Eqs.~(\ref{kk})-(\ref{WFkk}), the generic solution is given by the set in (\ref{sol2bis}), where in this case the coefficients $a(\kk)$ and $b(\kk)$ should fulfill the following set of linear equations:
\begin{eqnarray}\label{sol2ter}
{f}_{\kk_\circ}(z^b(k),k) &=& 
\int d\kk 
\, a(\kk) \, {f}_\kk(z^b(k),k)
\nonumber \\
&+&
\int d\kk \,
b(\kk) \, {f}_\kk(-z^b(k),k)
\ .
\end{eqnarray}
As previously stressed, the choice of the coefficients --and thus of the solution in (\ref{sol2bis})-- is  not unique, and can be parameterized according to Eq.~(\ref{c}).
The simplest choice $
a(\kk) = \delta(\kk-\kk_\circ)$ and $b(\kk) = 0$, 
implying   $c = 0$ in Eq.~(\ref{c}), corresponds the scattering-state solution ${f}_{\kk_\circ}(z,k)$.
Similarly, the choice $a(\kk) = b(\kk)$ implies $c = 1$ and, in the presence of a symmetric potential ($V(z) = V(-z)$), always provides a spatially symmetric solution, regardless of the profile of the boundary values.

\begin{figure}
\centering
\includegraphics*[width=8cm]{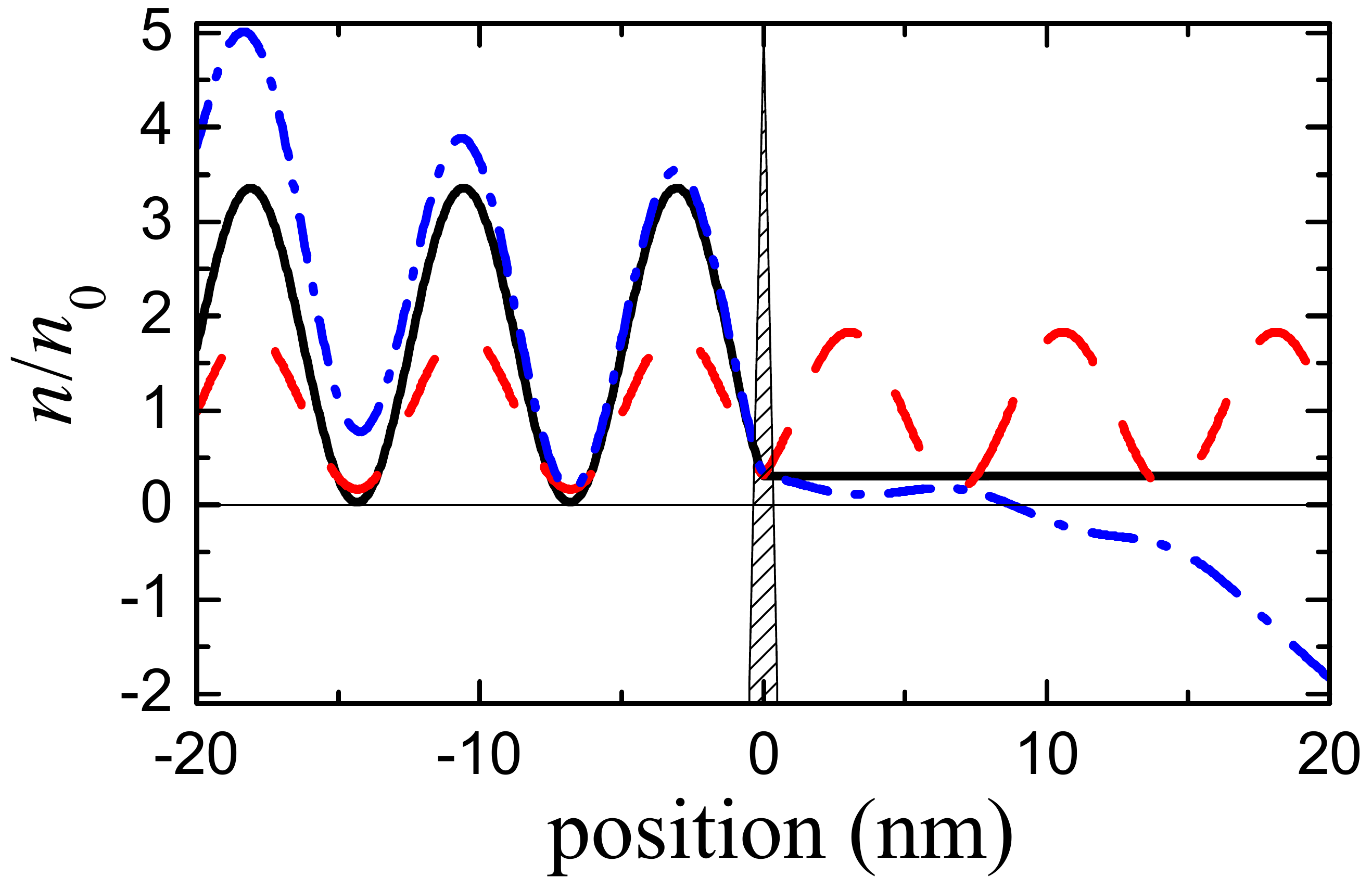}
\caption[]{(Color online) 
Non-uniqueness of the solution of the Wigner problem for the case of the delta-like potential barrier in (\ref{delta}). 
Three different spatial carrier-density profiles (see Eq.~(\ref{nzbis})) corresponding to the solution set in (\ref{sol2bis}):   the left scattering-state solution in (\ref{nzss2}) corresponding to $c = 0$ (black solid curve), the spatially symmetric solution corresponding to $c = 1$ (red dashed curve), and an unphysical solution (i.e., non-positive-definite) corresponding to $c = 0.05$ (blue dash-dotted curve) (see text).
The device parameters are $l = 40$\,nm, $\epsilon = 100$\,meV, and $\lambda = 1.5$, corresponding to a transmission coefficient $T \simeq 0.3$.}
\label{fig3}       
\end{figure}

The non-uniqueness of the solution is illustrated in Fig.~\ref{fig3}, which shows three carrier-density profiles corresponding to three different solutions of the same Wigner problem, namely of the Wigner transport equation (\ref{WECLSS}) applied to the delta-barrier potential (\ref{delta}) in the presence of the spatial boundary conditions corresponding to the left-state Wigner function in (\ref{WFss2}).
As expected, for $c = 0$ (black solid curve) one obtains the left-state density in (\ref{nzss2}), while for $c = 1$ (red dashed curve)  one deals with a spatially symmetric density. Moreover, for the intermediate value $c = 0.05$ (blue dash-dotted curve) we deal with an unphysical solution characterized by negative carrier-density values.
As already discussed in Sec.~\ref{s-U}, the presence of such unphysical solutions is not necessarily ascribed to the non-uniqueness discussed so far; indeed, also in the presence of energy dissipation --for which the solution is always unique (see below)--  one may easily obtain unphysical solutions by imposing arbitrary boundary conditions according to the conventional scheme of the semiclassical theory (see Fig.~\ref{fig8}); this feature, already pointed out in Ref.~[\onlinecite{Taj06}], appears to be the most severe limitation of conventional Wigner-function treatments.

Since in the presence of symmetric potentials any spatially symmetric discretization scheme (applied to the differential equation (\ref{WECLSS}) as well as to the Neumann-series expansion of Eq.~(\ref{WE-int})) 
returns spatially symmetric Wigner functions only,\cite{Taj06} 
one is forced to conclude that, among the infinite set of coefficients compatible with the given boundary conditions, such treatments automatically provide/select the symmetric choice $c = 1$.
This implies that, by applying such numerical treatments to the case of the delta-like potential (\ref{delta}) and using as boundary conditions the ones corresponding to the left-state Wigner function (\ref{WFss2}), 
one is expected to obtain the spatially-symmetric ($c = 1$) carrier density (red dashed curve) reported in Fig.~\ref{fig3}.
In order to validate this  conclusion, we have performed a numerical solution of the integral version of the Wigner equation (\ref{WE-int}) via a standard finite-difference technique. 
As expected, the result of our calculation (not reported here) is symmetric, and --apart from small deviations due to discretization as well as to phase-space cut-offs-- it coincides with the symmetric ($c = 1$) carrier density in Fig.~\ref{fig3}.\\

\section{Including energy-dissipation and decoherence phenomena}\label{s-IERDP}

In this section we shall discuss how to extend the coherent-limit treatment considered so far in order to account for energy-dissipation as well as decoherence phenomena induced by non-elastic scattering processes. It is worth noticing here that, because the Wigner-function equation~(\ref{WE2}) is obtained as the Weyl-Wigner transform of the density-matrix equation (\ref{SBE-op}), the reliability  of the Wigner approach crucially relies on the  degree of accuracy of such density-matrix formalism, which is in turn intimately related to the choice of the scattering superoperator $\Gamma$ in (\ref{SBE-op}). In particular, the preservation of the positive character of $\hat\rho$ by the dissipation/decoherence term in the density matrix equation (and hence of the Wigner Equation) is in fact a general  problem. 
Indeed, oversimplified approaches accounting for $\Gamma$ in a phenomenological way  or via kinetic treatments based on the conventional Markov limit\cite{Iotti05b} may lead to the violation of the positive-definite character of the density-matrix operator $\hat{\rho}$, and therefore to unphysical conclusions. 
Recently an alternative Markov procedure has been   proposed~\cite{Taj09b}  to overcome this serious limitation, showing that it is possible to derive a Lindblad-like scattering superoperator\cite{Lindblad76} of the form
\begin{equation}\label{Lindblad}
\Gamma \,(\hat{\rho})
=
\sum_s \left(
\hat A^s \hat\rho \hat A^{s \dagger}
- 
{1 \over 2} \left\{\hat A^{s \dagger} \hat A^s, \hat\rho\right\} 
\right) \, .
\end{equation}
In the low density limit, for each single-particle interaction mechanism $s$  one is thus able to perform a fully microscopic derivation of a corresponding Lindblad superoperator,\cite{Lindblad76}  
thereby preserving the positive-definite character of the density matrix $\hat\rho$ as well as of the corresponding Wigner function in (\ref{WF-op}).
However, the Wigner-function evolution induced by the Lindblad-like term in (\ref{Lindblad}) is definitely non-local within the phase-space ($\mathbf{r},\mathbf{k}$). \\

In order to adopt a local description and to apply the boundary-condition scheme discussed in Sec.~\ref{s-U}, a quite customary approach  is  the well known relaxation-time approximation~\cite{Rossi11}, which amounts to replacing the microscopic scattering superoperator (\ref{Lindblad}) with the relaxation-time term
\begin{equation}\label{RTA1}
\Gamma \,(\hat{\rho}) = 
-{
\hat{\rho} - \hat{\rho}^\circ
\over
{\tau}
}\ .
\end{equation}
Here
\begin{equation}\label{RTA2}
\hat{\rho}^\circ = \left({\Omega \over 2\pi}\right)^3\,
\int d\mathbf{k}\,
| \mathbf{k} \rangle 
f^\circ(\epsilon(\mathbf{k})-\mu^\circ) 
\langle \mathbf{k} | 
\end{equation}
is the equilibrium density-matrix operator (expressed via the corresponding Fermi-Dirac distribution $f^\circ$ characterized by a chemical potential $\mu^\circ$) and ${\tau}$ denotes a phenomenological (or macroscopic) relaxation time. 
By applying the Weyl-Wigner transform (\ref{WF-op}) to the scattering superoperator (\ref{RTA1}), one obtains the relaxation-time term  (\ref{RTAWF}) for the Wigner Equation, where the equilibrium Wigner function coincides with the Fermi-Dirac distribution $f^\circ$, i.e., $f^\circ(\mathbf{r},\mathbf{k}) = f^\circ(\epsilon(\mathbf{k})-\mu^\circ)$, and the Wigner-function relaxation time $\tau$ coincides with the relaxation time~${\tau}$ appearing in the density-matrix equation. The latter can be regarded as a sort of effective or average coherence time, and is mainly determined by carrier-phonon as well as carrier-carrier scattering. 

A quite customary way to generalize Eq.~(\ref{RTA1})   amounts to replace the parameter $\tau$ by a suitable operator $\hat{\tau}$, in order to account for possible space and/or momentum dependence of the relaxation time. However, such seemingly straightforward generalization may have non-trivial implications.  
Indeed, due to the non-local character of the Weyl-Wigner transform, such replacement in (\ref{RTA1}) gives rise to a non-local contribution in (\ref{WE2Gamma}), which is not simply given by the local term (\ref{RTAWF}), where the parameter $\tau$ is replaced by a space- and momentum-dependent relaxation time $\tau(\mathbf{r},\mathbf{k})$. 
Furthermore, while  a constant relaxation time ensures the positive-definite character of the single-particle density matrix --and therefore of the corresponding Wigner function--, such positive character does not hold for an arbitrary $\tau(\mathbf{r},\mathbf{k})$. For these reasons we shall consider ${\tau}$ as space and momentum independent.

Here, for the sake of simplicity and for similarity with previous works on the Wigner equation, we shall work within the relaxation time approximation, and consider the effect of dissipative/decoherence phenomena on the 
one-dimensional case previously analyzed in the coherent limit. The  relaxation-time approximation implies the appearance of  an additional contribution to the steady-state Wigner equation (\ref{WECLSS}), leading to the generalized Wigner transport equation
\begin{eqnarray}
v(k)\,{\partial f(z,k) \over \partial z} &=& - \int dk' {\cal
V}(z,k-k') f(z,k')
-\nonumber \\
& &   - {
f(z,k) -f^\circ(z,k)
\over
{\tau}
} \label{WERTAbis}
\ ,
\end{eqnarray}
with $f^\circ(z,k) \equiv f^\circ(\epsilon(k)-\mu^\circ)$ denoting the equilibrium Fermi-Dirac distribution induced by the host material.

In contrast to its coherent version in (\ref{WECLSS}), the above transport equation does not exhibit the $k \to -k$ symmetry discussed in Sec.~\ref{ss-NU}. From a physical point of view, the inclusion of this relaxation term destroys the time-reversal symmetry ($k \to -k$) responsible for the non-uniqueness previously discussed; it follows that, regardless of the value of the relaxation time ${\tau}$, the solution of the generalized Wigner transport equation (\ref{WERTAbis}) (compatible with given spatial boundary conditions) is always unique.

\begin{figure}
\centering
\includegraphics*[width=8cm]{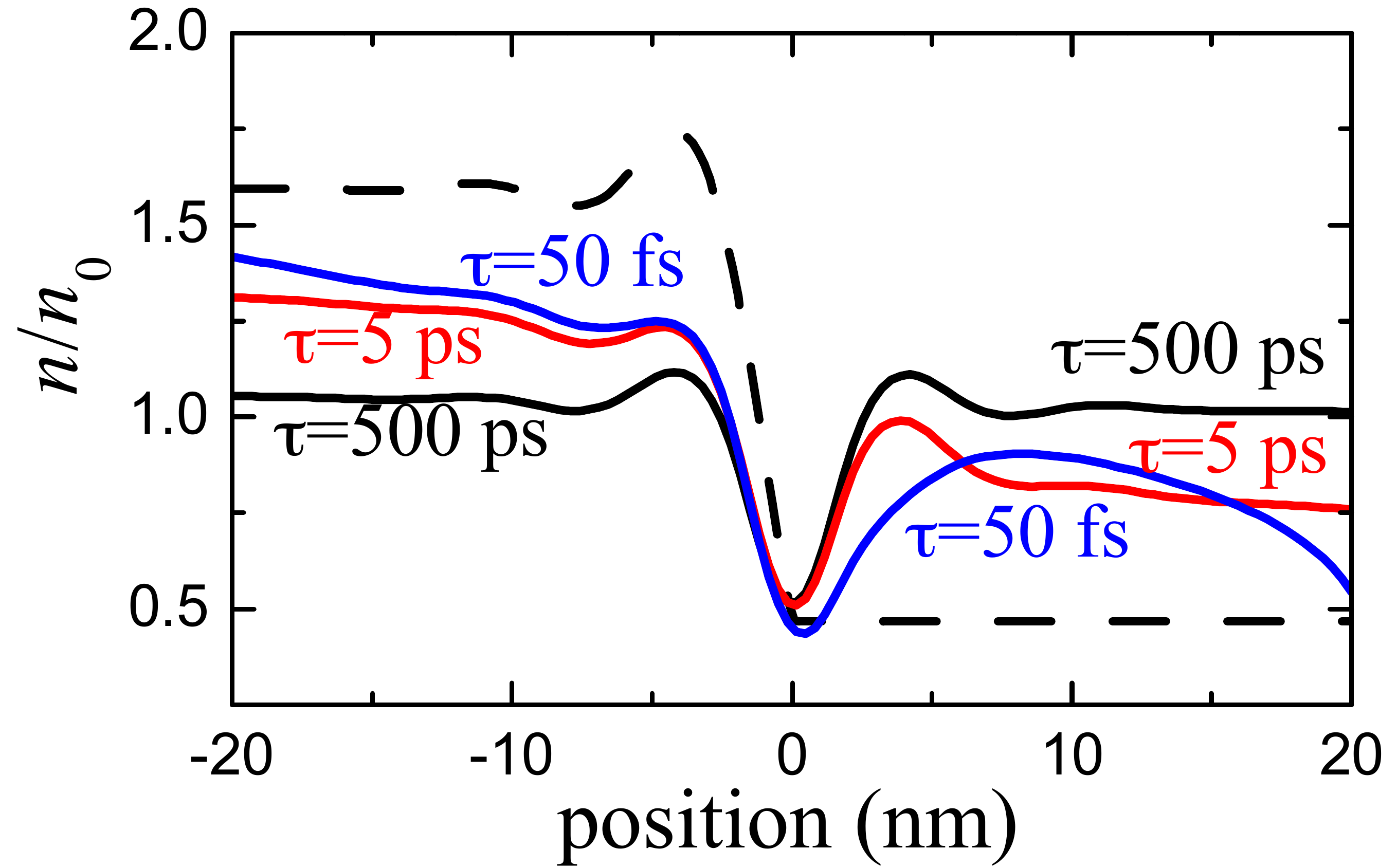}
\caption[]{
(Color online) 
Spatial carrier density for the case of the delta-like potential in (\ref{delta}).  
Comparison between the  result obtained by an analytical approach to the problem  (dashed curve) and via a numerical solution of the generalized Wigner equation (\ref{WERTAbis}), (based on a standard phase-space discretization scheme in terms of a $120 \times 120$ uniform grid) 
for different values of the relaxation time ${\tau}$ (solid curves), in the presence of a room-temperature carrier injection only from the left ($\mu_R \to -\infty$) for the same device and simulation parameters considered in Fig.~\ref{fig3} and for $\mu_L = 4 k_B T$
(see text).
}
\label{fig4}       
\end{figure}

In order to study the interplay between coherence and dissipation/decoherence, we have investigated carrier transport through the delta-barrier potential (\ref{delta}) in the presence of a quasiequilibrium thermal injection from the external reservoirs corresponding to the spatial boundaries
\begin{eqnarray}\label{thinj}
f\left(-{l \over 2}, k > 0\right) &=& f^\circ(\epsilon(k)-\mu_L)\nonumber \\
f\left(+{l \over 2}, k < 0\right) &=& f^\circ(\epsilon(k)-\mu_R)\ ,
\end{eqnarray}
where $\mu_L$ and $\mu_R$ denote the chemical potentials of the left and right reservoirs, and $f^\circ$ the corresponding quasiequilibrium Fermi functions.
In particular, we have considered the case of a room-temperature carrier injection from the left reservoir only ($\mu_R \to -\infty$).  

While in the coherent limit (${\tau} \to \infty$) such transport problem can be treated analytically via the Landauer-B\"uttiker formalism,\cite{Datta97} 
a numerical   solution of the Wigner equation (\ref{WERTAbis}) has been performed for different values of the relaxation time ${\tau}$. To this end we have chosen the same device and simulation parameters considered in Fig.~\ref{fig3}, and we have set $\mu_L = 4 k_B T$.  \\
Figure \ref{fig4} shows   the obtained spatial carrier density. 
The dashed curve  represents the coherent-transport result provided by a scattering-state calculation, and is therefore immune from the unphysical behaviors of the Wigner treatment pointed out above; explicitly, the density is  given by the thermal average of the pure-state carrier density in (\ref{nzss2}). In this coherent regime, one recovers the spatial density profile predicted by the Landauer-B\"uttiker theory:\cite{Datta97} while on the right hand side of the barrier the density is given by the fraction  of the carriers injected from the left reservoir that is transmitted across the barrier, on the left hand side the density accumulation is determined by the fraction of carriers reflected back to the left reservoir. Notice that, with respect to the pure-state density profile in Eq.~(\ref{nzss2}) (corresponding to a monoenergetic injection), here the thermal average leads to effective transmission and reflection coefficients; besides, the oscillatory contribution in Eq.~(\ref{nzss2}) averages out far from $z=0$. \\

\begin{figure}
\centering
\includegraphics*[width=8cm]{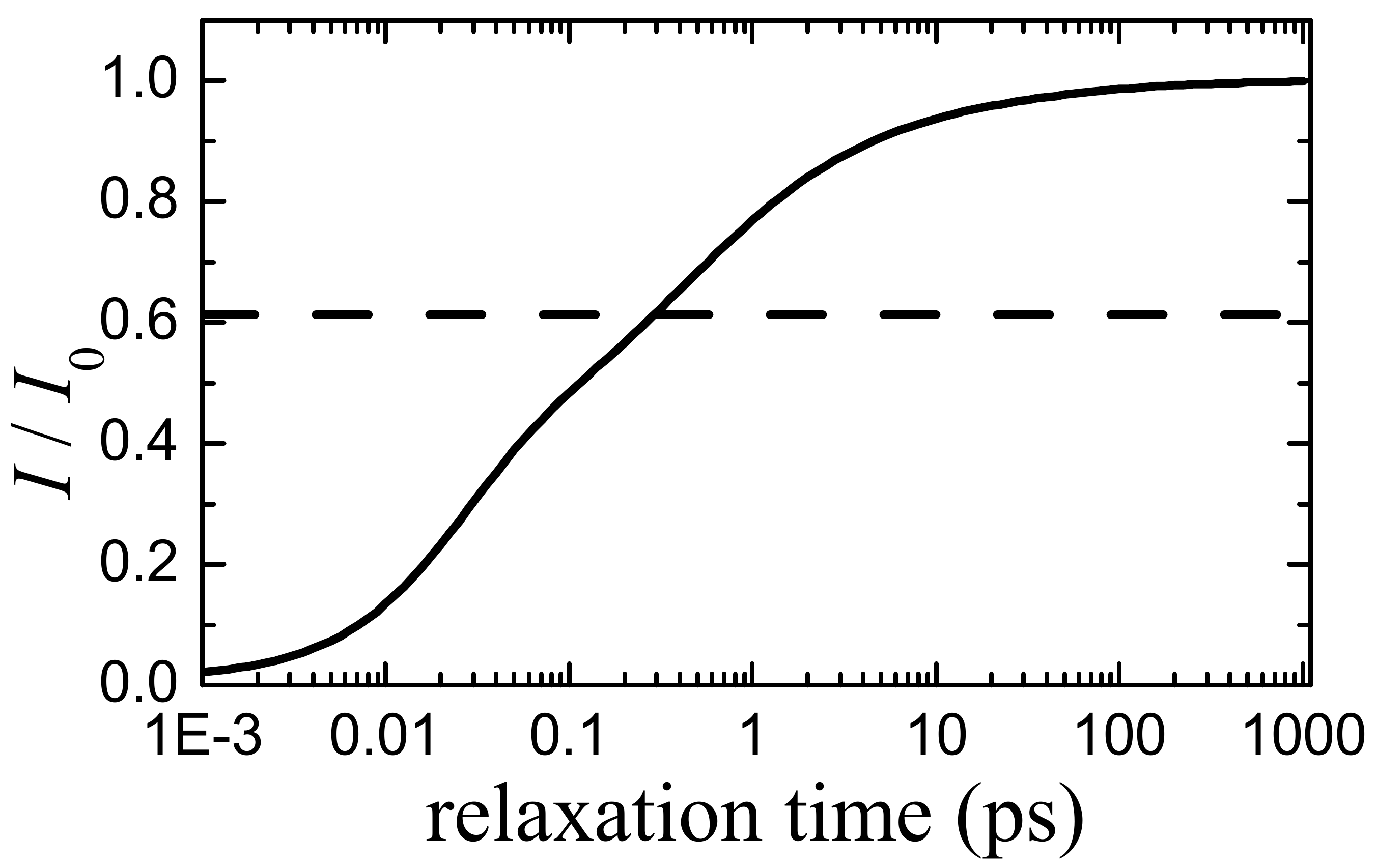}
\caption[]{
Charge current $I$ as a function of the relaxation time ${\tau}$ (in units of its potential- and dissipation-free value $I_\circ$) (solid curve) compared to the coherent-limit (${\tau} \to \infty$, dashed curve) current corresponding to the analytical charge density predicted by the Landauer-B\"uttiker theory (see dashed curve in Fig.~\ref{fig4})
(see text).
}
\label{fig5}       
\end{figure}

The solid curves in Fig.~\ref{fig4} correspond to three different values of the relaxation time: ${\tau} = 500$\,ps, ${\tau} = 5$\,ps, and ${\tau} = 50$\,fs. These timescales have to be compared to the average transit time, which is given by the ratio between the device length and the dissipation-free carrier drift velocity, and is of the order of $100$ fs. Thus, for ${\tau} = 500$\,ps the impact of energy relaxation and decoherence is expected to be definitely negligible, and  the coherent limit previously considered   should be recovered.
However, the charge density obtained via a numerical solution of the generalized Wigner equation (\ref{WERTAbis}) turns out to be significantly different from the dashed-curve one. Here,  in spite of the presence of the potential barrier, all injected carriers are transmitted (see below), and no reflection takes place. 
In contrast, for smaller values of the relaxation time the impact of dissipation and decoherence becomes significant: a decrease of ${\tau}$ (see solid curves in Fig.~\ref{fig4}) leads to a progressive decrease of the current, as shown in Fig.~\ref{fig5} (solid curve), which corresponds to an effective reflection of the injected carriers back to the left reservoir, induced by the relaxation-time term in (\ref{WERTAbis}).

The coherence-versus-dissipation scenario described so far is fully confirmed by the electronic-current analysis presented in Fig.~\ref{fig5}: here, we report the current $I$ as a function of the relaxation time ${\tau}$, in units of its potential- and dissipation-free value $I_\circ$, (solid curve) compared to the dissipation-free current predicted by the Landauer-B\"uttiker theory (dashed curve) 
corresponding to the dashed-curve charge-density profile in Fig.~\ref{fig4}.
As anticipated, in the coherent limit (${\tau} \to \infty$), while within the Landauer-B\"uttiker formalism the presence of the potential barrier leads to a significant attenuation of the current (${I /I_\circ} \simeq 0.6$, dashed curve), 
the dissipation-free current obtained via the Wigner equation (\ref{WERTAbis}) coincides with its potential-free value $I_\circ$ (see below).
For decreasing values of ${\tau}$ --corresponding to an increased impact of energy-dissipation/decoherence processes-- one observes a progressive reduction of the current (solid curve). 

From the numerical analysis reported so far, one  concludes that the conventional Wigner-function treatment leads, in general, to an overestimation of tunneling-like phenomena; such overestimation, particularly severe in the coherent-transport limit, may be quantitatively mitigated by the presence of non-elastic scattering processes. It is however important to point out that, from a fundamental point of view, the conventional Wigner-function treatment of coherent transport is intrinsically incompatible with well established results of the Landauer-B\"uttiker formalism.\cite{Datta97}

\begin{figure}
\centering
\includegraphics*[width=8cm]{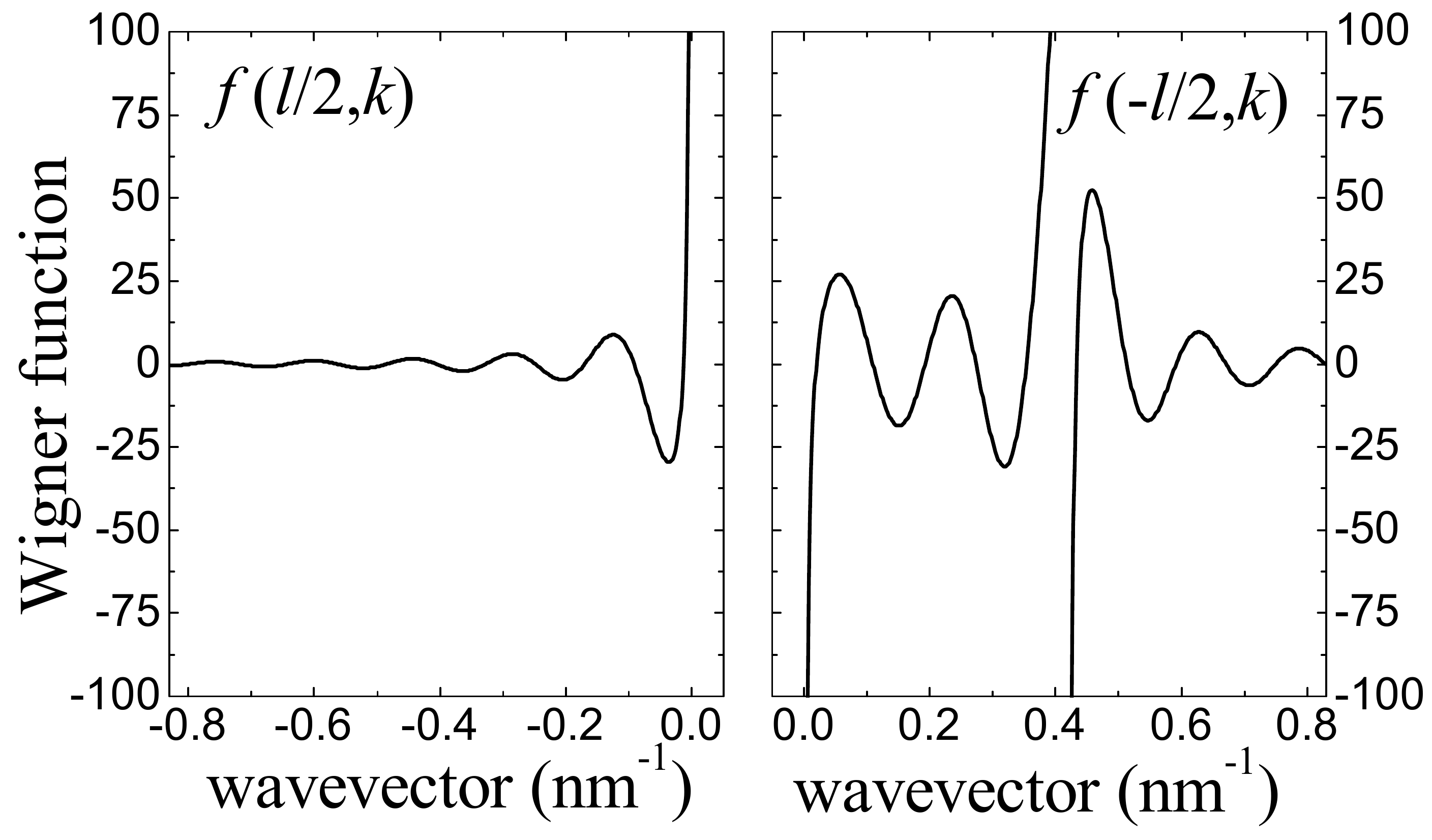}
\caption[]{
The inflow boundary profile determined by the analytical Wigner function (\ref{WFss2}) for a delta-like potential barrier (\ref{delta}), namely $f(z=l/2,k)$ for $k<0$ (left panel), and $f(z=-l/2,k)$ for $k>0$ (right panel). The Wigner function is plotted   in units of $2\pi/\Omega$, and the device parameters are the same as in Fig.~\ref{fig3}; in particular, the scattering-state wavevector is $\kk \simeq 4.2$\,nm$^{-1}$.
}
\label{fig6}       
\end{figure}

Different features of the Wigner-function treatment may induce the anomalous coherent-limit behavior reported in Figs.~\ref{fig4} and \ref{fig5}.
The first issue to be discussed is the validity of the thermal-injection boundary scheme in (\ref{thinj}).
Indeed, as recently pointed out in Refs.~[\onlinecite{Jiang10,Jiang11,Savio11}], such a semiclassical treatment/description of the boundary function $f^b(k)$ seems to be not necessarily compatible with the quantum-mechanical nature of a genuine Wigner function, as confirmed by the highly non-classical (i.e., non positive-definite) shape of the boundary conditions corresponding, e.g., to the scattering state solution (\ref{WFss2}).
This is clearly shown in Fig.~\ref{fig6}, where we report the left ($k > 0$) and right ($k < 0$) quantum-mechanical inflow boundary profile corresponding to the analytical Wigner function in (\ref{WFss2}); as anticipated, opposite to the usual semiclassical treatment, here the boundary function --corresponding to a left-scattering state of incoming wavevector $\kk \simeq 4.2$\,nm$^{-1}$-- involves all $k$ values and, more important, is not positive-definite.

We emphasize that, thanks to the presence of external carrier reservoirs in thermal or quasi-thermal equilibrium, the Wigner function of the quantum-device electron is expected to be far from a pure state. For this reason, in order to better compare the rigorous shape of the inflowing Wigner function with the usual Fermi-Dirac distribution of the semiclassical theory (employed in the conventional Wigner-function modeling), let us consider the Wigner function corresponding to a mixed state. To this aim, a thermal average is performed as incoherent superposition of the density matrices 
$| \kk \rangle \langle \kk |$ corresponding to the left- and right-scattering states in (\ref{ss+-}). In more explicit terms, this amounts to assume a density-matrix operator of the form
\begin{equation}\label{rhoth}
\hat{\rho} = {\Omega \over 2\pi}\,
\int \!d\kk  \, | \kk \rangle 
f^\circ_\kk \, 
\langle \kk |\ , 
\end{equation}
where the function
\begin{equation}\label{fcirckk}
f^\circ_\kk = \left\{ \begin{array}{lcl}
f^\circ(\epsilon(\kk)-\mu_L)
 & \mbox{for} & \quad  \kk > 0  \\ & & \\
f^\circ(\epsilon(\kk)-\mu_R)
 & \mbox{for} & \quad  \kk  < 0 
\end{array} \right.
\end{equation}
encodes the carrier distribution of the left and right reservoirs according to the sign of $\kk$.

By applying to the mixed/thermal-state density matrix (\ref{rhoth}) the one-dimensional version of the Weyl-Wigner transform in (\ref{WF-op}), the corresponding Wigner function comes out to be
\begin{equation}\label{WFth}
f(z,k) = {\Omega \over 2\pi}\,
\int \!d\kk  \,  {f}_\kk(z,k)\, f^\circ_\kk\ .
\end{equation}

\begin{figure}
\centering
\includegraphics*[width=8cm]{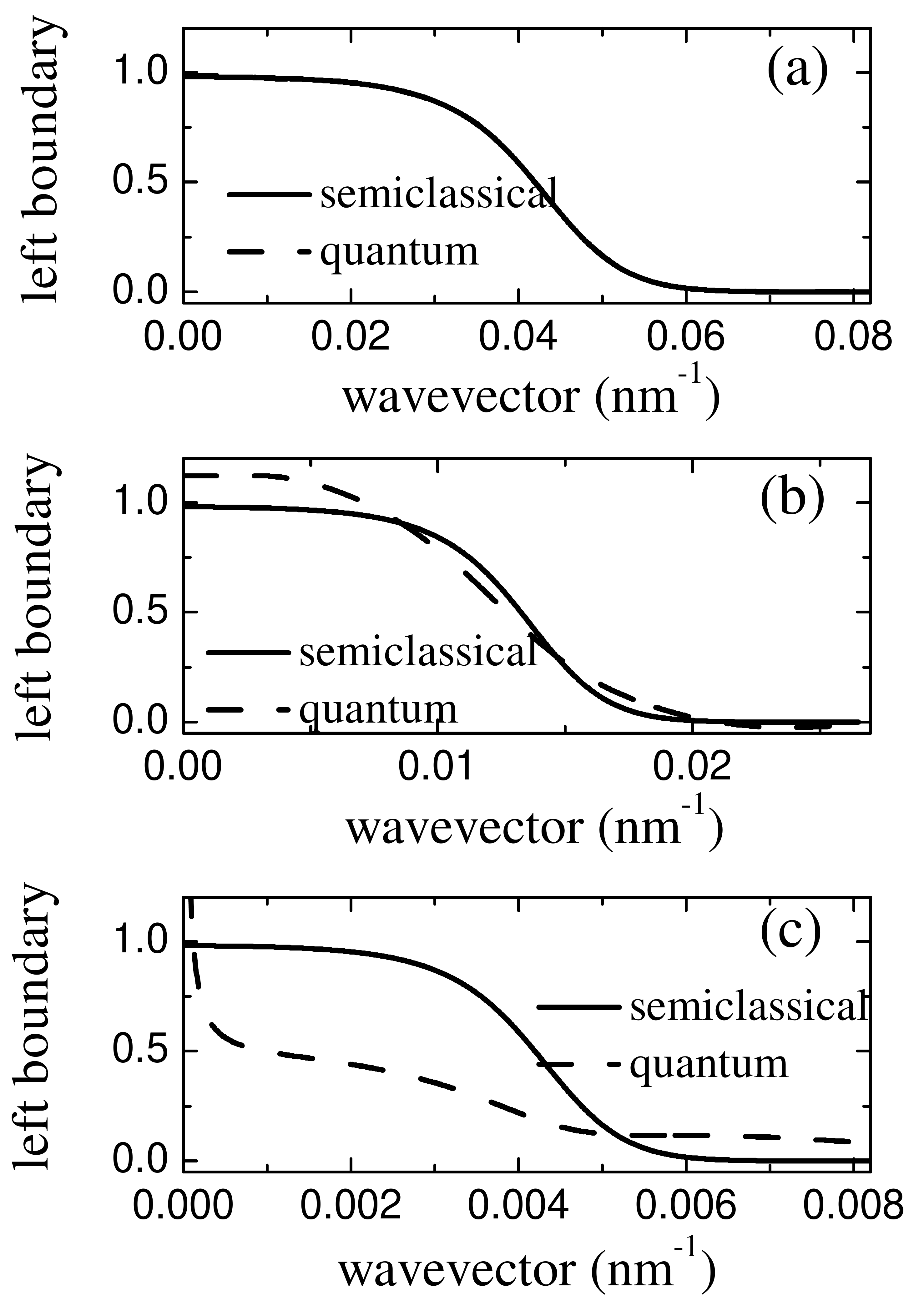}
\caption[]{
Case of the delta-like   potential barrier in (\ref{delta}). The value of the thermally averaged Wigner function in (\ref{WFth}) at the left boundary is plotted as a function of the wave vector $k$  (dashed curves) and is compared to  the semiclassical assumption of a  Fermi-Dirac distribution (solid curves), for three different temperature values:
$T = 300$\,K (a), $T = 30$\,K (b), and $T = 3$\,K (c). 
Here, the device parameters  are the same as in Fig.~\ref{fig3}, and for all three cases we have assumed a chemical potential $\mu_L = 4 k_B T$
(see text).
}
\label{fig7}       
\end{figure}

In order to quantify the impact of the above thermal average (with respect to the pure-state result in Fig.~\ref{fig6}), we have evaluated the inflowing part ($k > 0$) of the Wigner function (\ref{WFth}) at $z = -{l/2}$ (left boundary) for the same delta-like potential profile, assuming carrier injection from left only ($\mu_R \to -\infty$).
Figure \ref{fig7} shows a comparison between the left-contact Wigner function in (\ref{WFth}) (dashed curves) and the corresponding Fermi-Dirac distribution (solid curves) at three different temperatures:
$T = 300$\,K (a), $T = 30$\,K (b), and $T = 3$\,K (c); for all three cases we have assumed a chemical potential $\mu_L = 4 k_B T$.
As one can see, while at room temperature [panel (a)] the two curves coincide over a large range of $k$ values, for low temperatures [panels (b) and (c)] the value of the Wigner function on the left boundary significantly differs from the Fermi-Dirac distribution, unambiguous proof of the failure of a classical-like boundary condition treatment in the low-temperature limit. Such limitation was already pointed out by Frensley in its original paper,\cite{Frensley86} where he noticed that for the case of a resonant-tunneling diode the Wigner-function calculation resembles the experimental results at $T = 300$\,K,
but at lower temperatures it seriously underestimates the peak-to-valley ratio. 

From the boundary-condition analysis of Fig.~\ref{fig7} it follows that at room temperature the classical-like injection model in (\ref{thinj}) is definitely appropriate;
this seems to suggest that the anomalous coherent-limit results reported in Fig.~\ref{fig5} are the hallmark of a more general limitation of the whole Wigner-function transport modeling.
Indeed the electric current flowing through a generic quantum device, expressed in terms of the Wigner function $f(z,k)$ as
\begin{equation}\label{I1}
I(z) \propto \int_{-\infty}^{+\infty} v(k) f(z,k) dk  
\end{equation}
fulfills the charge continuity equation~\cite{note-CCE,note-CCEbis}. Because in steady-state conditions and in the absence of energy dissipation the current is $z$-independent [$I(z) = I_\circ$], it can be computed at any space point. In particular by evaluating Eq.~(\ref{I1}) at   the left boundary $z = -{l/2}$, and by splitting the integration domain into negative and positive $k$ values, one obtains
\begin{equation}\label{I2}
I_\circ \propto \int_{-\infty}^{0} v(k) f\left(-{l \over 2},k\right) dk\,+\,
\int_{0}^{+\infty} v(k) f\left(-{l \over 2},k\right) dk\ .
\end{equation}
For the case of a symmetric potential, it is possible to show that the (unique) solution of the generalized Wigner equation (\ref{WERTAbis}) in the coherent limit ${\tau} \to \infty$ is always spatially symmetric: $f(z,k) = f(-z,k)$.
Recalling that, at $z = -{l / 2}$ for $k > 0$ and at $z = +{l / 2}$ for $k < 0$, the Wigner function coincides with the inflow boundary function $f^b(k)$, 
and using the space symmetry of the Wigner function (between $-{l / 2}$ and $+{l / 2}$), Eq.~(\ref{I2}) can simply be rewritten as:
\begin{equation}\label{I3}
I_\circ \propto \int_{-\infty}^{+\infty} v(k) f^b(k) dk\ .
\end{equation}
This equation indicates that for the case of a symmetric potential the coherent-limit electric current is determined by the boundary values only, and is fully independent of the shape of the device potential profile. In particular, this leads to the  unphysical result that the value of the current turns out to be the same for a potential-free ballistic device as well as  for an infinitely high potential barrier. 

\begin{figure}
\centering
\includegraphics*[width=8cm]{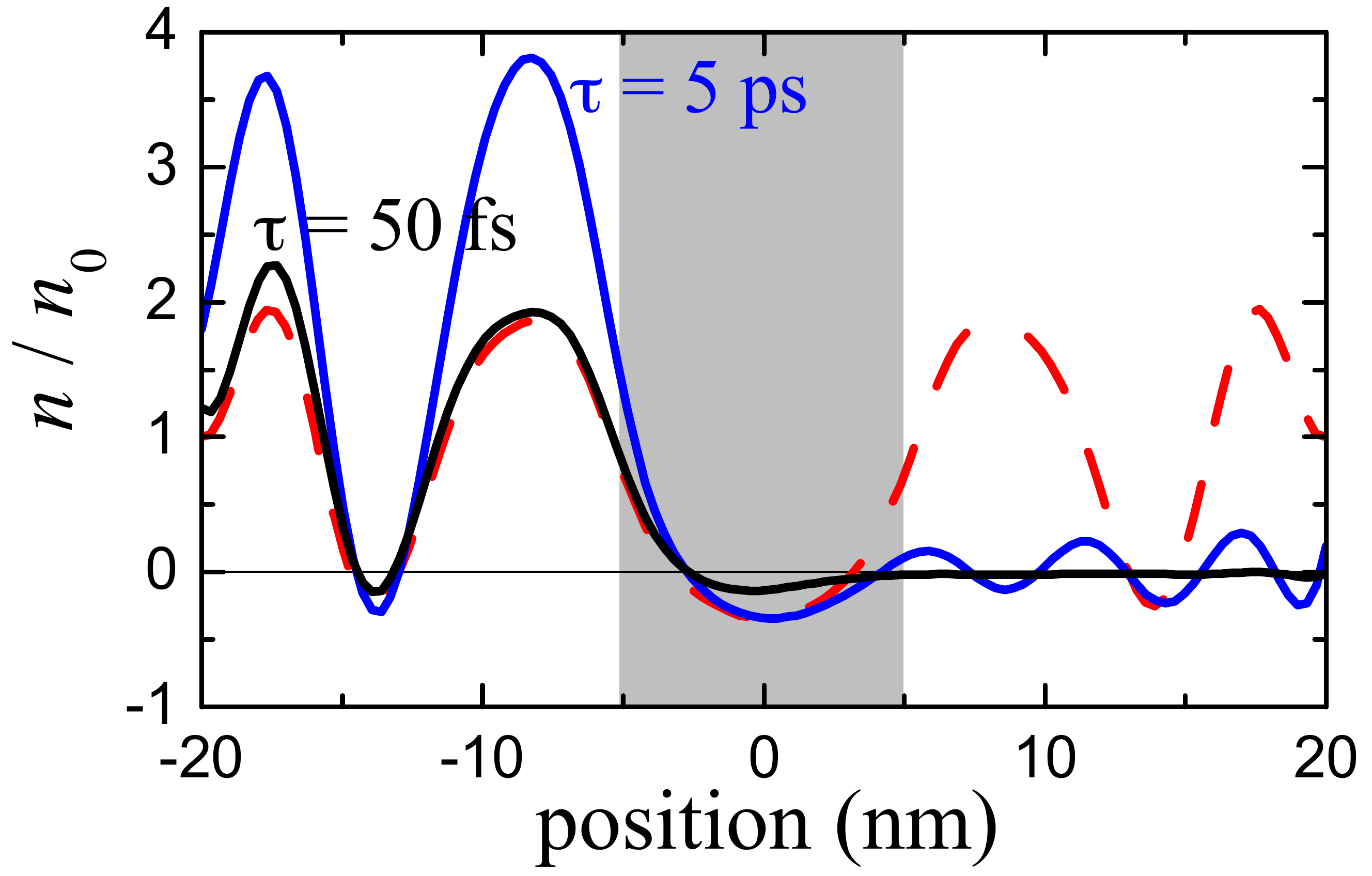}
\caption[]{(Color online) 
Spatial carrier density $n$ (in units of its barrier-free value $n_0$) corresponding to the rectangular-barrier profile in (\ref{fbp}) (device length $l = 40$\,nm, barrier width $a = 10$\,nm, and barrier height $V_0 = 150$\,meV) in the presence of a monoenergetic injection from the left ($\epsilon = 50$\,meV). 
Here the coherent-limit (${\tau}\to\infty$) density profile (dashed curve) is compared to the results corresponding to two different values of the relaxation time:
${\tau} = 5$\,ps and ${\tau} = 50$\,fs (solid curves). 
}
\label{fig8}       
\end{figure}

The coherence-versus-dissipation analysis presented so far may lead to conclude that, while in the coherent limit the Wigner-function modeling is highly problematic, in the presence of a significant energy-dissipation dynamics the results  are always physically acceptable. However, this is not the case. Indeed, one can consider the following situation: (i) replace the ideal  delta-like barrier in~(\ref{delta})  with a  more realistic  rectangular barrier with finite width $a$ and height $V_0$, i.e.,
\begin{equation}\label{fbp}
V(z) = V_0\, \theta\left({a \over 2}-|z|\right)\ ,
\end{equation}
and (ii) replace the thermal injection in (\ref{thinj}) with a simple monoenergetic carrier injection from left, i.e.,
\begin{equation}\label{mono}
f^b(k) \propto \delta(k-\kk) \ .
\end{equation}
The resulting spatial carrier-density profiles corresponding to the coherent limit (${\tau}\to\infty$) (dashed curve) as well as to two different values of the relaxation time ${\tau}$ (solid curves) are reported in Fig.~\ref{fig8}. 
As one can see, while energy dissipation induces once again a  spatial asymmetry (see also Fig.~\ref{fig4}), in the presence of a monoenergetic injection [see Eq.~(\ref{mono})] all  three density profiles display   negative-value regions. Thus   unphysical features also appear in the presence of a strong energy-dissipation dynamics.
This result is qualitatively similar to the one reported in [\onlinecite{Taj06}] for the case of a cosine-like potential, thus confirming the  physical limitations of the conventional Wigner-function modeling.

\section{Summary and conclusions}\label{s-SC}

In this work we have pointed out and explained some intrinsic limitations of the conventional quantum-device modeling strategy based on the well-known Wigner-function formalism. More specifically, we have provided a   definite answer to a few open questions related to the application of the conventional space-boundary condition scheme to the Wigner transport equation. By  combining analytical and numerical results, our  investigation  has shown that (i) in the coherent limit the solution of the Wigner equation (compatible with given boundary conditions) is not unique, and (ii) when decoherent/dissipative phenomena are taken into account within the relaxation-time approximation  the solution, although unique,  may be unphysical. Indeed it is not necessarily a physical Wigner function (see Fig.~\ref{fig8}), i.e., a Weyl-Wigner transform of a single-particle density matrix.

From a physical point of view, such intrinsic limitations of the standard (i.e., semiclassical) boundary condition scheme applied to the Wigner transport equation can be summarized as follows:
The essentially wrong ingredient in the conventional treatment is the artificial space separation between device active region ($|z| < {l / 2}$) and
external reservoirs ($|z| > {l / 2}$) (see Fig.~\ref{fig2}). Indeed, the latter is intrinsically incompatible with the well-known non-local character of quantum mechanics.

Our numerical results show that the above limitations are particularly severe in the coherent limit and/or in the presence of nonequilibrium carrier injection from the external reservoirs (e.g., monoenergetic distributions); this may explain why such anomalous behaviors are usually not experienced in conventional quantum-device modeling, since the latter is typically based on quasi-thermal injection  in the presence of a significant energy-dissipation dynamics.
In this respect some of the limitations discussed in this article may in principle also affect other modeling strategies based, e.g., on the non-equilibrium Green's functions.\cite{note-GF}
Indeed, in a recent study\cite{Jiang11} it has been shown that when the electric contacts are far enough from the device active region, the results of the inflow Wigner-function scheme and of conventional Green's function treatments coincide.
Since the anomalous coherent-limit behavior reported in Figs.~\ref{fig4} and \ref{fig5} is not related to the boundary location, it seems that such limitation may also affect Green's function treatments; however, in order to provide a definite answer to this point, a specific investigation is needed.

In order to overcome the basic limitations of the Wigner-function modeling discussed in this article, the crucial step could be to replace the local (i.e., classical-like) boundary condition-scheme treatment of the device-reservoir interaction with a fully non-local approach; to this end, in order to ensure/maintain the positive-definite character of the electronic density matrix, a possible strategy is to describe the system/device-environment/reservoir interaction via a Lindblad-like coupling term.\cite{Lindblad76} This task is beyond the purpose of the present work, and is discussed elsewhere.\cite{OQD12}


\appendix

\section{Analytical evaluation of the Wigner function for a delta-like potential profile}

The goal of this Appendix is twofold:
on the one hand, we shall discuss the analytical derivation of the Wigner function corresponding to the delta-like potential in (\ref{delta}); on the other hand, we shall verify that such Wigner function fulfils the corresponding Wigner equation.

We start by introducing the general prescription for the analytical evaluation of the one-dimensional pure-state Wigner function in (\ref{WFphi}). 
To this end, we shall limit ourselves to quantum-mechanical states whose wavefunctions have   different analytical expressions on the left ($L$) and on the right ($R$) of the space-coordinate origin ($z = 0$), i.e., 
\begin{equation}\label{philr}
\phi(z) =
\cases{
\phi_L(z) & {\rm for} \quad $z < 0$ \cr
\phi_R(z) & {\rm for} \quad $z > 0$
} \ .
 \end{equation}
This applies to any potential profile of the form
\begin{equation}
V(z) = \Lambda \delta(z) + V_\circ \theta(z)\ ,
\end{equation}
which includes, as particular cases, the delta-like potential in (\ref{delta}) as well as the   step-potential (not considered in this work).

In order to evaluate the explicit form of the Wigner function in (\ref{WFphi}), the key step is to perform the integration over $z'$; to this end, for any given value $z$ the arguments of the two wavefunctions may assume negative (left) as well as positive (right) values according to the value of $z'$. In particular one obtains
\begin{eqnarray}\label{t1}
z' < -2z &\quad\rightarrow\quad& z+{z' \over 2} < 0 \nonumber \\
z' > -2z &\quad\rightarrow\quad& z+{z' \over 2} > 0 \nonumber \\
z' > 2z &\quad\rightarrow\quad& z - {z' \over 2} < 0 \nonumber \\
z' < 2z &\quad\rightarrow\quad& z - {z' \over 2} > 0 \ .
\end{eqnarray}
According to the above set of inequalities, the integration domain in (\ref{WFphi}) 
($-\infty < z' < +\infty$) needs to be split into three different subdomains.
More specifically, for $z> 0$ we have
\begin{eqnarray}\label{split1}
f(z,k) &=& \int_{-\infty}^{+\infty} dz' e^{-i kz'}\, \phi\left(z+{z' \over 2}\right) \phi^*\left(z-{z' \over 2}\right)  \nonumber \\
&=& \int_{-\infty}^{-2z} dz' e^{-i kz'}\, \phi_L\left(z+{z' \over 2}\right) \phi^*_R\left(z-{z' \over 2}\right)  \nonumber \\
&+& \int_{-2z}^{+2z} dz' e^{-i kz'}\, \phi_R\left(z+{z' \over 2}\right) \phi^*_R\left(z-{z' \over 2}\right)  \nonumber \\
&+& \int_{+2z}^{+\infty} dz' e^{-i kz'}\, \phi_R\left(z+{z' \over 2}\right) \phi^*_L\left(z-{z' \over 2}\right) \ ,\nonumber \\
\end{eqnarray}
while for $z < 0$ we have
\begin{eqnarray}\label{split2}
f(z,k) &=& \int_{-\infty}^{+\infty} dz' e^{-i kz'}\, \phi\left(z+{z' \over 2}\right) \phi^*\left(z-{z' \over 2}\right)  \nonumber \\
&=& \int_{-\infty}^{+2z} dz' e^{-i kz'}\, \phi_L\left(z+{z' \over 2}\right) \phi^*_R\left(z-{z' \over 2}\right)  \nonumber \\
&+& \int_{+2z}^{-2z} dz' e^{-i kz'}\, \phi_L\left(z+{z' \over 2}\right) \phi^*_L\left(z-{z' \over 2}\right)  \nonumber \\
&+& \int_{-2z}^{+\infty} dz' e^{-i kz'}\, \phi_R\left(z+{z' \over 2}\right) \phi^*_L\left(z-{z' \over 2}\right) \ .\nonumber \\
\end{eqnarray}
Taking into account that for both cases ($z > 0$ and $z < 0$) the last integral is exactly the complex conjugate of the first one, i.e.,
\begin{displaymath}
\int_{-\infty}^{-2|z|} dz' e^{-i kz'}\, \phi_L\left(z+{z' \over 2}\right) \phi^*_R\left(z-{z' \over 2}\right)
\end{displaymath}
\begin{equation}\label{deltastep}
= \left(\int_{+2|z|}^{+\infty} dz' e^{-i kz'}\, \phi_R\left(z+{z' \over 2}\right) \phi^*_L\left(z-{z' \over 2}\right)\right)^*\ ,
\end{equation}
and that
\begin{equation}
\int_{2|z|}^{\infty} f(z') dz' = \int_0^\infty f(z') dz' - \int_0^{2|z|} f(z') dz' \ ,
\end{equation}
the two results in (\ref{split1}) and (\ref{split2}) can be combined as:
\begin{eqnarray}\label{split3}
f( z , k ) \!&=&\! 2\Re\left(\int_0^\infty dz' e^{-i kz'}\, \phi_R\left(z\!+\!{z' \over 2}\right) \phi^*_L\left(z\!-\!{z' \over 2}\right)\right) \nonumber \\
&-&\! 2\Re\left(\int_0^{2|z|} dz' e^{-i kz'}\, \phi_R\left(z\!+\!{z' \over 2}\right) \phi^*_L\left(z\!-\!{z' \over 2}\right)\right) \nonumber \\
&+& \!\theta(z) \int_{-2|z|}^{2|z|} dz' e^{-i kz'}\, \phi_R\left(z\!+\!{z' \over 2}\right) \phi^*_R\left(z\!-\!{z' \over 2}\right)  \nonumber \\
&+& \!\theta(-z) \int_{-2|z|}^{2|z|} dz' e^{-i kz'}\, \phi_L\left(z\!+\!{z' \over 2}\right) \phi^*_L\left(z\!-\!{z' \over 2}\right)  \ .\nonumber \\
\end{eqnarray}
The above prescription can be easily extended to any piece-wise-constant potential, like, e.g., multi-step as well as multi-barrier profiles. 

For the particular case of the delta-like potential profile (\ref{delta}), the explicit form of the left ($z < 0$) and right ($z > 0$) part ($\phi_L$ and $\phi_R$) of the electron wavefunction is provided by the scattering states in (\ref{ss+-}).
In particular, by inserting into Eq.~(\ref{split3}) the explicit form of the left scattering state (i.e., $\kk > 0$), after a lengthy but straightforward calculation one obtains the Wigner function in Eq.(\ref{WFss2}).
It is possible to show that the Wigner function corresponding to the right scattering state ($\kk < 0$) can simply be obtained from the left-scattering one in (\ref{WFss2}) by replacing $z$ with $-z$ as well as $k$ with $-k$: ${f}_{-\kk}(z,k) = {f}_\kk(-z,-k)$. To this aim, we observe that the application to Eq.~(\ref{WFphi}) of the Wigner-space transformation $z,k \to -z,-k$ is equivalent to replacing $\phi^{ }(z)$ with $\phi^*(-z)$, the very same wavefunction transformation linking left and right scattering states. 

As a final step, let us verify that the Wigner function (\ref{WFss2}) is a solution of the corresponding Wigner equation. 
By inserting  the potential superoperator (\ref{calVdelta}) corresponding to the delta-like barrier profile (\ref{delta}) into Eq.(\ref{WECLSS}), the explicit form of the Wigner equation comes out to be 
\begin{equation}\label{WECLSSdelta}
v(k)\,{\partial f(z,k) \over \partial z} = {4 \Lambda \over 2\pi \hbar} \int dk' \sin\left(2(k-k') z \right) f(z,k')\ .
\end{equation}
In order to verify that the Wigner function (\ref{WFss2}) is indeed a solution of the above Wigner transport equation, let us now evaluate separately its kinetic and potential terms. 
As far as the kinetic contribution is concerned, after a tedious but straightforward calculation one gets
\begin{widetext}
\begin{equation}\label{Kss}
v(k)\,{\partial f(z,k) \over \partial z} 
= 
- {4 \lambda \hbar\kk \over \Omega \, m^* \left(1+\lambda^2\right)}
\,
\left(
\sin\left(2(\kk-k)z\right) 
- \theta(-z) \lambda \left(
\cos\left(2(\kk+k)z\right)
-
\cos\left(2(\kk-k)z\right)
\right)
\right)\ .
\end{equation}
\end{widetext}
Let us now come to the potential contribution in (\ref{WECLSSdelta}). By inserting the explicit form of the scattering state Wigner function (\ref{WFss2}), again after a tedious but straightforward calculation one gets
\begin{widetext}
\begin{equation}\label{Vss}
{4 \Lambda \over 2\pi \hbar} \int dk' \sin\left(2(k\!-k') z \right) f(z,k')
= 
- {4 \Lambda \over \Omega \hbar \left(1+\lambda^2\right)}
\left(
\sin\left(2(\kk-k)z\right)
- \theta(-z) \lambda
\left(
\cos\left(2(\kk+k)z\right)
-
\cos\left(2(\kk-k)z\right)
\right)
\right)\ .
\end{equation}
\end{widetext}
By inserting the explicit forms of the kinetic and potential terms in (\ref{Kss}) and (\ref{Vss}) into the Wigner transport equation (\ref{WECLSSdelta}), we clearly see that the left-state Wigner function (\ref{WFss2}) is indeed a solution of the Wigner transport equation for $\lambda_\kk = {m^* \Lambda / \hbar^2|\kk|}$, the very same prescription in (\ref{lambda}) obtained via a direct solution of the Schr\"odinger equation.

\begin{acknowledgments}

We are extremely grateful to Carlo Jacoboni and David Taj for stimulating and fruitful discussions. F.D. acknowledges support from FIRB 2012 project ÓHybridNanoDevÓ (Grant No.RBFR1236VV).

\end{acknowledgments}


\begin{thebibliography}{99}
\bibitem{note-degeneracy}
This is due to the time-reversal symmetry which, in turn, is intimately related to the energetic double degeneracy of the scattering states.
Indeed, the wavefunction of any quantum-mechanical state with negative energy ($\epsilon < 0$) is always real, and the time-reversal symmetry just mentioned plays no role; it follows that in the presence of localized states only --e.g., parabolic potentials or quantum wells with infinite-height barriers-- the solution of the Wigner transport equation (\ref{WECLSS}) is expected to be unique, in agreement with the conclusions of the analysis presented in Ref.~[\onlinecite{Barletti01}].
\bibitem{note-normalizz} 
Just like for planewaves, one requires that $\int_{-\Omega/2}^{+\Omega/2} |\phi^{\pm}(z)|^2 \, dz =1$ for $\Omega \rightarrow \infty$.
\bibitem{note-CCE}
In order to show that in steady-state conditions and in the absence of energy dissipation the electric current in (\ref{I1}) is $z$-independent, it is enough to show that its space derivative, 
\begin{equation}\label{I4}
{d I(z) \over dz} \propto \int dk v(k)\,{\partial f(z,k) \over \partial z}\ ,
\end{equation}
in the coherent limit is always equal to zero. To this end, recalling that in steady-state conditions [see Eq.~(\ref{WECLSS})]
\begin{equation}\label{I5}
v(k)\,{\partial f(z,k) \over \partial z} = - \int dk' {\cal
V}(z,k-k') f(z,k')\ ,
\end{equation}
the space derivative in (\ref{I4}) can also be expressed as:
\begin{equation}\label{I6}
{d I(z) \over dz} \propto 
\int dk \int dk' {\cal V}(z,k-k') f(z,k')\ .
\end{equation}
By setting $k'' = k - k'$ and recalling the antisymmetric character of the potential superoperator (${\cal V}(z,-k'') = - {\cal V}(z,k'')$), we finally obtain
\begin{equation}\label{I7}
{d I(z) \over dz} \propto 
\int dk' f(z,k') \int dk'' {\cal V}(z,k'') = 0\ .
\end{equation}
\bibitem{note-CCEbis}
Importantly, within the relaxation-time approximation (see Eq.~(\ref{WERTAbis})) the charge continuity equation just recalled is not valid anymore, and the steady-state current (see Fig.~\ref{fig5}) is typically evaluated averaging $I(z)$ over the device active region.
\end{thebibliography}

\end{document}